%% file: main.tex
\documentclass[preprint,10pt,authoryear]{elsarticle}
\usepackage{graphicx, verbatim, url, amssymb, amsmath, amsfonts}

\journal{arXiv}

\begin{document}
\begin{frontmatter}

\title{Effective and efficient approximations of the generalized inverse of the graph Laplacian matrix with an application to current-flow betweenness centrality}

\author{Enrico Bozzo}
\address{Department of Mathematics and Computer Science, University of Udine, Italy}

\author{Massimo Franceschet}
\address{Department of Mathematics and Computer Science, University of Udine, Italy}

\begin{abstract}
We devise methods for finding approximations of the generalized inverse of the graph Laplacian matrix, which arises in many graph-theoretic applications. Finding this matrix in its entirety involves solving a matrix inversion problem, which is resource demanding in terms of consumed time and memory and hence impractical whenever the graph is relatively large. Our approximations use only few eigenpairs of the Laplacian matrix and are parametric with respect to this number, so that the user can compromise between effectiveness and efficiency of the approximated solution. We apply the devised approximations to the problem of computing current-flow betweenness centrality on a graph. However, given the generality of the Laplacian matrix, many other applications can be sought. We experimentally demonstrate that the approximations are effective already with a constant number of eigenpairs. These few eigenpairs can be stored with a linear amount of memory in the number of nodes of the graph and, in the realistic case of sparse networks, they can be efficiently computed using one of the many methods for retrieving few eigenpairs of sparse matrices that abound in the literature. 

\end{abstract}


\end{frontmatter}

\input{introduction.tex}

\input{laplacian.tex}

\input{approximations.tex}
\input{complexity.tex}
\input{centrality.tex}
\input{experiments.tex}

\input{conclusion.tex}

\bibliographystyle{elsarticle-harv}
\bibliography{bibliometrics}

\end{document}

%% file: introduction.tex
\section{Introduction} \label{introduction}

The graph Laplacian is an important matrix that can tell us much about graph structure. It places on the diagonal the degrees of the graph nodes and elsewhere information about the distribution of edges among nodes in the graph. The graph Laplacian matrix, as well as its Moore-Penrose generalized inverse \citep{BIG03}, are useful tools in network science that turn up in many different places, including random walks on networks, resistor networks, resistance distance among nodes, node centrality measures, graph partitioning, and network connectivity \citep{GBS08,N10}. In particular, among measures of centrality of graph nodes, betweenness quantifies the extent to which a node lies between other nodes. Nodes with high betweenness are crucial actors of the network since they control over information (or over whatever else flows on the network) passing between others. Moreover, removal from the network of these brokers might seriously disrupt communications between other vertices \citep{N04b,BF05}.

The computation of the generalized inverse of the Laplacian matrix is demanding in terms of consumed time and space and thus it is not feasible on relatively large networks. On the other hand, there are today large databases from which real networks can be constructed, including technological, information, social, and biological networks \citep{BE05,NBW06,N10}. These networks are voluminous and grow in time as more data are acquired. We have, therefore, the problem of running computationally heavy algorithms over large networks. The solution investigated in the present work is the use of approximation methods: algorithms that compute a solution near to the exact one and, as a compromise, that run using much less resources than the exact algorithm. 

We propose a couple of approximation methods to compute the generalized inverse of the Laplacian matrix of a graph. Both methods are based on the computation of few eigenpairs (eigenvalues and the corresponding eigenvectors) of the Laplacian matrix \citep{GM10}, where the number of computed eigenpairs is a parameter of the algorithm. The first method, called cutoff approximation, uses the computed eigenpairs in a suitable way for the approximation of the actual entries of the generalized inverse matrix. The second method, named stretch approximation, takes advantage of both the computed eigenpairs as well as of an estimation of the excluded ones. Both approximation methods can be applied to estimate current-flow betweenness centrality scores for the nodes of a graph. We experimentally show, using both random and scale-free network models, that the proposed approximation are both effective and efficient compared to the exact methods. In particular the stretch method allows to estimate, using a feasible amount of time and memory, a ranking of current-flow betweenness scores that strongly correlates with the exact ranking. 

The layout of the paper is as follows. Section \ref{laplacian} introduces the notions of Laplacian matrix and its Moore-Penrose generalized inverse and recalls some basic properties of these matrices. In Section \ref{approximations} we define cutoff and stretch approximations. Moreover, we theoretically show that stretch approximation is more effective than cutoff approximation. We review the methods for inverting a matrix and for finding few eigenpairs of a matrix, which are crucial operations in our contribution, in Section \ref{complexity}. Current-flow betweenness centrality is illustrated in Section \ref{centrality}. We formulate the definition in terms of the generalized inverse of the Laplacian matrix, which allows us to use cutoff and stretch approximations to estimate betweenness scores. A broad experimental analysis is proposed in Section \ref{experiments} in order to investigate  effectiveness and efficiency of the devised approximation methods. Section \ref{conclusion} concludes the paper.

%% file: laplacian.tex
\section{The graph Laplacian and its generalized inverse} \label{laplacian}

Let $\mathcal{G}=(V,E,w)$ be an undirected weighted graph with $V$ the set of nodes, $E$ the set of edges, and $w$ a vector such that $w_i > 0$ is the positive weight of edge $i$, for $i = 1, \ldots, |E|$. We denote by $n$ the number of nodes and $m$ the number of edges of the graph.  The weighted Laplacian of $\mathcal{G}$ is the symmetric matrix $$G=D-A$$ where $A$ is the weighted adjacency matrix of the graph and $D$ is the diagonal matrix of the generalized degrees (the sum of the weights of the incident arcs) of the nodes. 

In order to obtain more insight on the properties of the graph Laplacian it is useful to express the
matrix in another form. Let $B\in\mathbb{R}^{n\times m}$ be the incidence matrix of the graph such
that, if edge $l$ connects two arbitrarily ordered nodes $i$ and $j$ then $B_{i,l}=1$, $B_{j,l}=-1$,
while $B_{k,l}=0$ for $k\neq i,j$. Given a vector $v$, the square diagonal matrix whose diagonal
entries are the elements of $v$ is denoted with ${\rm Diag}(v)$. It holds that $G=B\:{\rm
Diag}(w)B^T$. Thus $G$ besides symmetric is positive semidefinite, so that it has real and
nonnegative eigenvalues that is useful to order $0\le\lambda_1\le\lambda_2\le\ldots\le \lambda_n$.
If $e$ denotes a vector of ones, then $De=Ae$ so that $Ge=0$. It follows that $\lambda_1=0$ is the
smallest eigenvalue of $G$. We will assume throughout the paper that $\mathcal{G}$ is connected. In
this case all other eigenvalues of $G$ are strictly positive \citep{GBS08}: $$0=\lambda_1<\lambda_2\le\ldots\le \lambda_n.$$

Since $\lambda_1 = 0$, the determinant of $G$ is null and hence $G$ cannot be inverted. As a
substitute for the inverse of $G$ we use the Moore-Penrose generalized inverse of $G$, that we  simply call generalized inverse of $G$ 
\citep{BIG03}. As customary, we denote this kind of generalized inverse with $G^+$.  
It is convenient
to define $G^+$ starting from the spectral decomposition of $G$. Actually, since $G$ is symmetric it
admits the spectral decomposition $$G=V\Lambda V^T$$ where $\Lambda={\rm
Diag}(0,\lambda_2,\ldots,\lambda_n)$ and the columns of $V$ are the eigenvectors of $G$. Notice that
$V$ is an orthogonal matrix, that is $VV^T=I=V^TV$. 

By using the spectral decomposition of $G$, its generalized inverse can be defined as
follows \begin{equation}\label{mpenrose} G^+=V{\rm
Diag}(0,\frac{1}{\lambda_2},\ldots,\frac{1}{\lambda_n}) V^T=\sum_{j=2}^n
\frac{1}{\lambda_j}V(:,j)V(:,j)^T, \end{equation}  where $V(:,j)$
(respectively, $V(j,:)$) denotes the $j$-th column (respectively, row) of matrix $V$.

Observe that $G^+$ inherits from $G$ the property of being symmetric and positive semidefinite.
Moreover, $G^{+}$ shares the same nullspace of $G$, as is true in general for the Moore-Penrose
generalized inverse of a symmetric matrix. Thus, since $Ge=0$, it turns out that $G^+e=0$. By
setting $J=ee^T$, it follows that $GJ=JG=G^{+}J=JG^{+}=O$, where $O$ is a matrix of all zeros. By
using the eigendecompositions of $G$ and $G^+$ it is easy to show that $(G+1/n J)(G^++1/n J)=I$. It
follows that 
\begin{equation} \label{inverse}
G^+=(G+ee^T/n)^{-1}-ee^T/n, 
\end{equation}
a formula that can be found in \citet{GBS08} and is
used implicitly in \citet{BF05}. Another useful consequence of the above equalities is
$$GG^+=G^{+}G=I-1/n J.$$

The generalized inverse of the graph Laplacian $G$ is useful to solve a linear system of the form $Gv=b$ for some known vector $b$, which arises in many applications. The range of a matrix $G$ is the linear space of vectors $b$ for which the
system $Gv=b$ has a solution. Since $G$ is symmetric its range is the space orthogonal to 
its nullspace. The nullspace of $G$ is one-dimensional and spanned by the vector $e$ with all components equal to unity.  Hence, the range of $G$ is made up by the vectors $x$ such that $e^Tx=0$, or, equivalently, that sum up
to zero. It follow that the linear system $Gv=b$ has solutions if $b$ sums up to zero. By the linearity, the difference of two solution belongs to the nullspace of $G$, and this implies that if we are able to find an arbitrary solution $v^*$, then
all the other solutions are of the form $v^*+\alpha e$, $\alpha\in\mathbb{R}$. 

As well known, $v^*=G^+b$ is the minimum Euclidean norm solution of the system $Gv=b$, i.e., it is 
the element having minimum Euclidean norm  in the affine space of the solutions  \citep{BIG03}. For
completeness, we notice that, no matter if  $b$ belongs to the range of $G$ or not, $v^*=G^+b$ is
the minimum Euclidean norm solution of the problem  $\min_x \|Gx-b\|_2$.

%% file: approximations.tex
\section{Approximations of the generalized inverse} \label{approximations}

In this section we propose two approximations of the the generalized inverse $G^+$ of the graph Laplacian matrix $G$. For $k=2,\ldots,n$, we define the $k$-th cutoff approximation of $G^+$ as: 

\begin{equation}\label{cutoff}
T^{(k)}=\sum_{j=2}^k\frac{1}{\lambda_j}V(:,j)V(:,j)^T.
\end{equation} 

In all computations, we never materialize matrix $T^{(k)}$, but we represent it with the $k-1$ eigenpairs that define it. This representation of $T^{(k)}$ can be stored using $O(k n)$ space, that is $O(n)$ if $k$ is constant. Moreover, computing an entry of $T^{(k)}$ using its eigenpair representation costs $O(k)$, that is $O(1)$ if $k$ is constant. 

As $k$ increases, the matrices $T^{(k)}$ are more and more accurate approximations of $G^+$. Actually $G^+ = T^{(n)}$ and, for $k<n$, 

\begin{equation}\label{additive}
G^+=T^{(k)}+\sum_{j=k+1}^n\frac{1}{\lambda_j}V(:,j)V(:,j)^T. 
\end{equation} 

It holds that $\|G^+-T^{(k)}\|_2\le\|G^+-M\|_2$ for every $M\in\mathbb{R}^{n\times n}$ having
rank less or equal to $k-1$ \citep{GLV96}. Moreover, for $k=2,\ldots,n-1$, the relative 2-norm error of the $k$-th cutoff approximation is:

$$\frac{\|G^+-T^{(k)}\|_2}{\|G^+\|_2}=\frac{1/\lambda_{k+1}}{1/\lambda_2}=\frac{\lambda_2}{\lambda_{
k+1}}.$$

The second approximation of $G^+$ exploits the following observation. If many of the excluded eigenvalues $\lambda_j$, for $j$ larger than $k$, are close to each other, we might approximate them with a suitable value $\sigma$. We define the $k$-th stretch approximation of $G^+$ as:

\begin{equation}\label{stretch}
S^{(k)}=T^{(k)}+\sum_{j=k+1}^n\frac{1}{\sigma}V(:,j)V(:,j)^T.
\end{equation}

It is worth observing that the use of $S^{(k)}$ does not involve any significant additional cost with respect to the use of $T^{(k)}$. Indeed, since $\sum_{j=1}^nV(:,j)V(:,j)^T = VV^T = I$, then 

\begin{eqnarray*}
S^{(k)}&=&T^{(k)}-\sum_{j=1}^k\frac{1}{\sigma}V(:,j)V(:,j)^T+\sum_{j=1}^n\frac{1}{\sigma}V(:,j)V(:,j
)^T\\ &=&
\frac{1}{\sigma}I-\frac{1}{\sigma}V(:,1)V(:,1)^T+\sum_{j=2}^k(\frac{1}{\lambda_j}-\frac{1}{\sigma})V
(:,j)V(:,j)^T. 
\end{eqnarray*}

Notice that the normalization of the eigenvector of $G$ associated with the eigenvalue $\lambda_1=0$ yields $V(:,1)= e / \sqrt{n}$, where $e$ is a vector with all components equal to unity. It follows that, knowing the value $\sigma$, the $k$-th stretch approximation $S^{(k)}$ can be represented using $k-1$ eigenpairs, hence the space needed to store the representation of $S^{(k)}$ and the time needed to compute its entries do not increase with respect to the use of $T^{(k)}$. 

On the other hand, the use of $S^{(k)}$ instead of $T^{(k)}$ allows to improve the bound on the approximation error. Actually, since 

$$G^+-S^{(k)}=G^+-T^{(k)}-\sum_{j=k+1}^n\frac{1}{\sigma}V(:,j)V(:,j)^T,$$ 

from Equation \ref{additive} we obtain

$$\|G^+-S^{(k)}\|_2=\|\sum_{j=k+1}^n(\frac{1}{\lambda_j}-\frac{1}{\sigma})V(:,j)V(:,j)^T
\|_2=\max_{j=k+1,\ldots,n}|\frac{1}{\lambda_j}-\frac{1}{\sigma}|.$$ 

Assuming, as reasonable, that $\lambda_{k+1}\leq \sigma\leq \lambda_n$, we have that

$$\frac{1}{\lambda_n}-\frac{1}{\lambda_j}\leq\frac{1}{\sigma}-\frac{1}{\lambda_j}\leq \frac{1}{\lambda_{k+1}}-\frac{1}{\lambda_j}$$ 

and hence 

$$\max_{j=k+1,\ldots,n}|\frac{1}{\lambda_j}-\frac{1}{\sigma}|\leq \frac{1}{\lambda_{k+1}}-\frac{1}{\lambda_n} = \gamma$$ 

so that 

$$\frac{\|G^+-S^{(k)}\|_2}{\|G^+\|_2} \le \lambda_2 \gamma < \frac{\lambda_2}{\lambda_{k+1}} = \frac{\|G^+-T^{(k)}\|_2}{\|G^+\|_2}.$$

Therefore, the relative 2-norm error of the stretch approximation $S^{(k)}$ is strictly less than the relative 2-norm error of the cutoff approximation $T^{(k)}$, as soon as we chose $\sigma$ within $\lambda_{k+1}$ and $\lambda_{n}$. Moreover, the closer $\lambda_{k+1}$ and $\lambda_n$, the better the stretch approximation. 

The optimal choice for $\sigma$, that is, the value that minimizes the 2-norm relative error, is the harmonic mean of $\lambda_{k+1}$ and $\lambda_n$:

$$\frac{1}{\sigma}=\frac{1}{2}\left(\frac{1}{\lambda_{k+1}}+\frac{1}{\lambda_n}\right)$$ 

With this choice we reduce of one half the bound of the approximation error: 

$$\max_{j=k+1,\ldots,n}|\frac{1}{\lambda_j}-\frac{1}{\sigma}|\le \frac{\gamma}{2}.$$ 

We have used this choice of $\sigma$ in all our experiments. Notice that the computation of $\sigma$
implies computing two additional eigenvalues, namely $\lambda_{k+1}$ and $\lambda_n$, but not the
corresponding eigenvectors. To avoid this additional cost, we might reasonably assume that
$\lambda_n$ is big so that its reciprocal is small, and that $\lambda_{k+1}$ is close to
$\lambda_{k}$, so that the optimal value of $\sigma$ is approximately $2 \lambda_{k}$. 

%% file: complexity.tex
\section{Methods for matrix inversion and for finding few eigenpairs} \label{complexity}

Finding the generalized inverse of the graph Laplacian matrix involves solving a matrix inversion
problem (Equation \ref{inverse}). Inverting a matrix is, however, computational demanding in terms
of used time and memory. Given a matrix $A$, the columns of $A^{-1}$ can be computed by solving the
linear systems $Ax=e_i$ for $i=1,\ldots,n$, where $e_i$ is the vector whose $i$-th entry is equal to
one and the other entries are equal to zero. If a direct method is used, then $A$ is factorized and the
factorization is used to solve the systems. This typically costs $O(n^3)$ floating point operations
and $O(n^2)$ memory locations (the inverse of a matrix is almost invariably dense even if the input matrix is sparse) \citep{AHU74}. In particular, the complexity of matrix product and
matrix inversion are the same, and the best known lower bound for matrix product, obtained for
bounded arithmetic circuits, is $\Omega(n^2\log n)$ \citep{R03}. If an iterative method is used
then, in the case where $A$ has a conditioning independent from the dimension, or a good
preconditioner can be found, the number of iterations becomes independent from the dimension.  Since
every iteration costs $O(m)$, then the cost is $O(m n)$ floating point operations and $O(n^2)$
memory locations to store the inverse. If the matrix is sparse, the number of operations is
quadratic. Otherwise the number of operations has to be multiplied by an additional factor dependent
from the conditioning of $A$. For an introduction to iterative methods to solve linear systems and
to preconditioning see \citet{S03}.

Instead of computing the entire generalized inverse of the Laplacian matrix, our approximation
methods compute and store only few eigenpairs of the Laplacian matrix.  If a matrix $A$ is big and
sparse, then the computation of few eigenpairs of $A$ can be made by means of iterative methods
whose basic building block is the product of $A$ by a vector, which has linear complexity if $A$ is
sparse. One of the simplest among these methods  is orthogonal iteration \citep{GLV96}, a
generalization of the power method. The method, while simple, can be quite slow since the number of
iterations depends on the distance between the sought eigenvalues of $A$ and experimental evidence
shows that the eigenvalues nearest to zero are clustered, in particular for sparse networks
\citep{ZCY10}.

On the other hand, one of the most widely used algorithms is the Lanczos method with implicit
restart, implemented by ARPACK \citep{LSY98}. This is the method we have used in our experiments.
For the computation of few smallest eigenpairs of a matrix $A$ the method works in the so called
shift and invert mode. In other words the Lanczos method is applied to $(A-\sigma I)^{-1}$ being
$\sigma$ a suitable shift. To do this, the matrix $A-\sigma I$ is factorized before the iteration
begins and the factorization is used to solve the sequence of linear systems that arises during the
calculation. This accelerates the method but the factors are surely much less sparse than the matrix
itself. This, combined with the clustering of the eigenvalues near zero, leads to a nonlinear
scaling, which was observed also in our experiments.

Alternative approaches are Jacobi-Davidson and Deflation Accelerated Conjugate Gradient
\citep{BP02}, that seem to be highly competitive with the Lanczos method. In particular in the
Jacobi-Davidson method it is still needed to solve inner linear systems, but the factorization is
avoided and substituted by the use of preconditioned iterative Krylov spaces based methods.
Deflation Accelerated Conjugate Gradient sequentially computes the eigenpairs by minimizing the
Rayleigh quotient $q(z)= z^T A z / z^Tz$ over the subspace orthogonal to the eigenvectors previously
computed. Finally, we mention the multilevel algorithm implemented in the HSL\_MC73 routine of the HSL
mathematical software library, that however only computes the second smallest eigenpair
\citep{HS03}.

%% file: centrality.tex
\section{Current-flow betweenness centrality} \label{centrality}

A large volume of research on networks has been devoted to the concept of centrality \citep{S66,F79,B05,N10}. This research addresses the fundamental question: \textit{Which are the most important or central vertices in a network?} There are four measures of centrality that are widely used in network analysis: degree centrality, eigenvector centrality, closeness, and  betweenness. Here, we focus on betweenness centrality. 

Betweenness measures the extent to which a node lies on paths between other nodes. Nodes with high betweenness might have considerable influence within a network by virtue of their control over information (or over whatever else flows on the network) passing between others. They are also the ones whose removal from the network will most disrupt communications between other vertices because they lie on the largest number of paths between other nodes. 

Typically, only geodesic paths are considered in the definition, obtaining a measure that is called shortest-path betweenness. Here, we study current-flow betweenness, which includes contributions of all paths, although longer paths give a lesser contribution \citep{N04b,BF05}. For a given node, current-flow betweenness measures the current flow that passes through the vertex when a unit of current is injected in a source node and removed from a target node, averaged over all source-target pairs. Equivalently, it is equal to the net number of times that a random walk on the graph passes through the node on its journey, averaged over a large number of trials of the random walk.

We next give the precise definition of current-flow betweenness centrality in terms of resistor networks.  Consider a network in which the edges are resistors and the nodes are junctions between resistors. Each edge is assigned with a positive weight indicating the conductance of the edge. The resistance of an edge is the inverse of its conductance. Outlets are particular nodes where current enters and leaves the network. A vector $u$ called supply defines them: a node $i$ such that $u_i \neq 0$ is an outlet; in particular, if $u_i > 0$ then node $i$ is a source and current enters the network through it, while if $u_i < 0$ then node $i$ is a target and current leaves the network through it. Since there should be as much current entering the network as leaving it, we have that $\sum_i u_i = 0$. We consider the case where a unit of current enters the network at a single source $s$ and leaves it at a single target $t$. That is, $u_{i}^{(s,t)} = 0$ for $i \neq s,t$, $u_{s}^{(s,t)} = 1$, and $u_{t}^{(s,t)} = -1$. We are interested in how current flows through the network, for an arbitrary choice of source and target outlets.

Let $v_{i}^{(s,t)}$ be the potential of node $i$, measured relative to any convenient reference potential, for source $s$ and target $t$ outlets. Kirchhoff's law of current conservation implies that the node potentials satisfy the following equation for every node $i$:

\begin{equation} \label{kirchhoff}
\sum_j A_{i,j} (v_{i}^{(s,t)} - v_{j}^{(s,t)}) = u_{i}^{(s,t)}
\end{equation}

where $A$ is the weighted adjacency matrix of the network. The current flow through edge $(i,j)$ is the quantity $A_{i,j} (v_{i}^{(s,t)} - v_{j}^{(s,t)})$, that is, the difference of potentials between the involved nodes multiplied by the conductance of the edge: a positive value indicates that the current flows in a direction (say from $i$ to $j$), and negative value means that the current flows in the opposite direction. Hence, Krichhoff's law states that the current flowing in or out of any node is zero, with the exception of the source and target nodes.

In matrix form, equation \ref{kirchhoff} reads:

\begin{equation} \label{kirchhoff2}
(D-A) v^{(s,t)} = Gv^{(s,t)} = u^{(s,t)}
\end{equation}

where $D$ is a diagonal matrix such that the $i$-th element of the diagonal is equal to $\sum_j A_{i,j}$, that is, it is the (generalized) degree of node $i$. Recall that $G = D-A$ is the graph Laplacian matrix. As noticed in Section \ref{laplacian}, if $G^+$ is the generalized inverse of the Laplacian matrix $G$, then the potential vector:

\begin{equation} \label{kirchhoff3}
v^{(s,t)} = G^+ u^{(s,t)}
\end{equation}

This means that the potential of node $i$ with respect to source $s$ and target $t$ outlets is given by $v_{i}^{(s,t)} = G^{+}_{i,s} - G^{+}_{i,t}$. Therefore, the generalized inverse matrix $G^+$ contains information to compute all node potentials for any pair of source-target nodes. 

\begin{figure}[t]
\begin{center}
\includegraphics[scale=0.40, angle=-90]{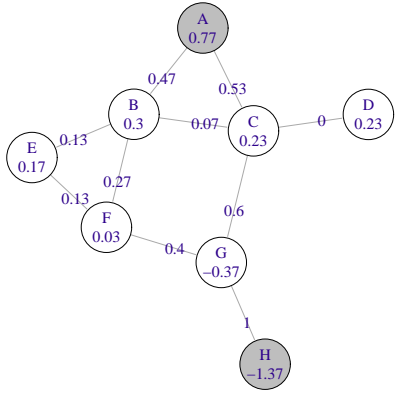}
\caption{A resistor network with all resistances equal to unity. Each node is identified with a letter and is labelled with the value of its potential when a unit current is injected in node A and removed from node H. Each edge is labelled with the absolute current flowing on it.}
\label{plot.resistor}
\end{center}
\end{figure}

An example of resistor network with node potential solution is provided in Figure \ref{plot.resistor}. Notice that Kirchhoff's law is satisfied for each node. For instance, the current entering in node B is 0.47 (from node A) which equals the current leaving node B, which is again 0.47 (0.13 to E, 0.27 to F, and 0.07 to C). Moreover, the current leaving the source node A is 1, and the current entering the target node H is also 1. Notice that there is no current on the edge from C to D, since both nodes have the same potential. Any other potential vector obtained from the given solution by adding a constant is also a solution, since the potential differences remain the same, and hence Kirchhoff's law is satisfied. The given potential vector is, however, the solution with minimum Euclidean norm.

We have now all the ingredients to define current-flow betweenness centrality. As observed above, given a source $s$ and a target $t$, the absolute current flow through edge $(i,j)$ is the quantity $A_{i,j} |v_{i}^{(s,t)} - v_{j}^{(s,t)}|$. By Kirchhoff's law the current that enters a node is equal to the current that leaves the node. Hence, the current flow $F_{i}^{(s,t)}$ through a node $i$ different from the source $s$ and a target $t$ is  half of the absolute flow on the edges incident in $i$:

\begin{equation} \label{flow}
F_{i}^{(s,t)} = \frac{1}{2} \sum_j A_{i,j} |v_{i}^{(s,t)} - v_{j}^{(s,t)}|
\end{equation}

\begin{figure}[t]
\begin{center}
\includegraphics[scale=0.40, angle=-90]{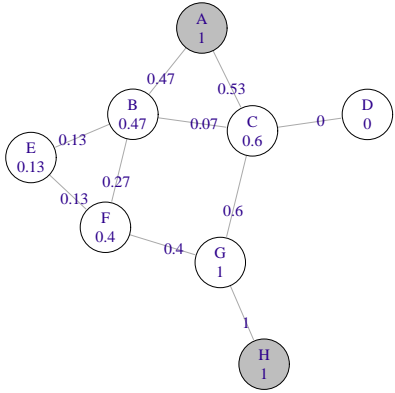}
\caption{A resistor network with all resistances equal to unity (this is the same network of Figure \ref{plot.resistor}). Each node is now labelled with the value of flow through it when a unit current is injected in node A and removed from node H. Each edge is labelled with the absolute current flowing on it.}
\label{plot.resistor2}
\end{center}
\end{figure}

Moreover, the current flows $F_{s}^{(s,t)}$ and $F_{t}^{(s,t)}$ through both $s$ and $t$ are set to 1, if end-points of a path are considered part of the path (this is our choice in the rest of this paper), or to 0 otherwise. Figure \ref{plot.resistor2} gives an example. Notice that the flow from A to H through node G is 1 (all paths from A to H pass eventually through G), the flow through F is 0.4 (a proper subset of the paths from A to H go through F and these paths are generally longer than for G), and the flow through E is 0.13 (a proper subset of the paths from A to H go through E and these paths are generally longer than for F)

Finally,  the current-flow betweenness centrality $b_i$ of node $i$ is the flow through $i$ averaged over all source-target pairs $(s,t)$:

\begin{equation} \label{cent}
b_i = \frac{\sum_{s < t} F_{i}^{(s,t)}}{(1/2) n (n-1)}
\end{equation}

Since the potential $v_{i}^{(s,t)} = G^{+}_{i,s} - G^{+}_{i,t}$, with $G^+$ the  generalized inverse of the graph Laplacian, Equation \ref{flow} can be expressed in terms of elements of $G^+$ as follows:

\begin{equation} \label{flow2}
\begin{array}{lcl}
F_{i}^{(s,t)} & = & \frac{1}{2} \sum_j A_{i,j} |G^{+}_{i,s} + G^{+}_{j,t} - G^{+}_{i,t} - G^{+}_{j,s}|
\end{array}
\end{equation}

Hence, if we replace in Equation \ref{flow2} matrix $G^+$ with its $k$-th cutoff approximation $T^{(k)}$ (or its $k$-th stretch approximation $S^{(k)}$), we get an approximated value of current-flow betweenness centrality for node $i$. 

The computational complexity of  Equation \ref{cent} is as follows. We denote with $k_i$ the number of neighbors of node $i$, that is, the number of edges incident in $i$. Assuming we have matrix $G^+$, computing Equation \ref{flow2} for given $i$, $s$, and $t$ costs $O(k_i)$, if $i$ has at least one neighbor, or $O(1)$ otherwise. Hence Equation \ref{cent} for a specific $i$ costs $O(n^2 \cdot \min(k_i, 1))$. Since $\sum_i \min(k_i, 1) = O(n+m)$, Equation \ref{cent} for all nodes costs $O(n^2 (n+m))$, that is, $O(n^3)$ if the graph is sparse.\footnote{This cost can be improved to $O(m n \log n)$, hence $O(n^2 \log n)$ for sparse networks,  as shown in \citet{BF05}.} Furthermore, the  complexity of computing the equation using either cutoff or stretched approximations increases of a factor of $k$, since an entry $(i,j)$ of $T^{(k)}$ or of $S^{(k)}$ can be obtained in $O(k)$. If $k$ is contant with respect to $n$, then there is no asymptotic increase of complexity. Clearly, if $n$ is large, the cost of Equation \ref{cent} is prohibitive.

A possible solution for the bottleneck of Equation \ref{cent} is the adoption of a probabilistic approach by computing Equation \ref{flow2} only for a sample of source-target pairs \citep{BF05}. If we choose at random $\alpha n$ source nodes, with $0 < \alpha < 1$, and, for each source, we choose at random $\alpha n$ target nodes, then the sample of source-target pairs has size $\alpha^2 n^2$. For instance, if $\alpha = 0.1$, then $\alpha^2 = 0.01$ and hence the cost of computing Equation \ref{cent} declines of two orders of magnitude.    

An alternative solution to avoid the bottleneck of Equation \ref{cent} is to use cutoff approximation with $k = 2$ (only the second smallest eigenpair is used). In this case Equation \ref{cent} significantly simplifies. Indeed, if $k = 2$, then $T^{(2)}_{i,j} = (1 / \lambda_2) V_{i,2} V_{j,2}^T$. It follows that flow $F^{(st)}_i$ can be approximated by:

$$
\begin{array}{lcl}
\hat{F}_{i}^{(s,t)}
& = & \frac{1}{2} \sum_j A_{i,j} |T^{(2)}_{i,s} + T^{(2)}_{j,t} - T^{(2)}_{i,t} - T^{(2)}_{j,s}| \\\\
& = & \frac{1}{2} \sum_j A_{i,j} \frac{1}{\lambda_2}|V_{s,2} - V_{t,2}| |V_{i,2} - V_{j,2}| \\\\
& = & \frac{1}{2\lambda_2} |V_{s,2} - V_{t,2}| \sum_j A_{i,j} |V_{i,2} - V_{j,2}| 
\end{array}
$$

Notice that the sum in the formula for the flow is now independent on the source-target pair.
Hence the approximated betweenness of $i$ is:

$$
\hat{b}_i = \frac{2}{n (n-1)} \sum_{s < t} \frac{1}{2\lambda_2} |V_{s,2} - V_{t,2}| \sum_j A_{i,j} |V_{i,2} - V_{j,2}| = C \sum_j A_{i,j} |V_{i,2} - V_{j,2}|
$$ 

where $C$ is a constant. This means that the approximated betweenness $\hat{b}_i$ is proportional to the quantity $\sum_j A_{i,j} |V_{i,2} - V_{j,2}|$. Notice that $\hat{b}_i$ is a local version of $b_i$ that depends only on the neighborhood of $i$. Ignoring the multiplicative constant $C$, the approximated betweenness $\hat{b}_i$ can be computed in $O(n+m)$ for all nodes, hence in linear time with respect to the size of the network.

\begin{figure}[t]
\begin{center}
\includegraphics[scale=0.40, angle=-90]{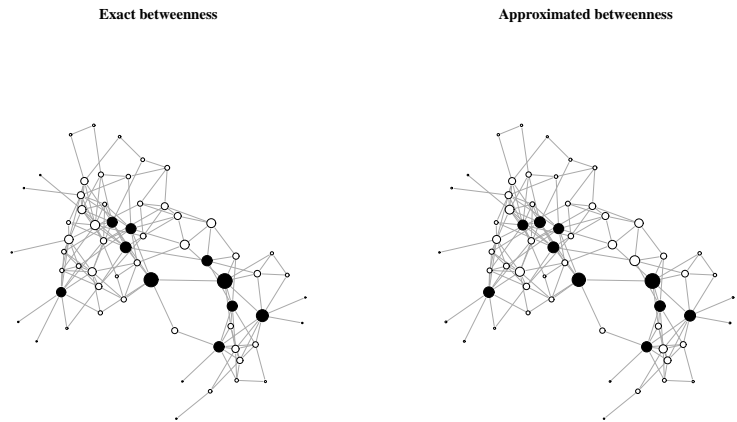}
\end{center}
\caption{A dolphin social network. The size of the nodes is proportional to the current-flow betweenness. The exact solution is shown on the left and the approximated one is shown on the right. Black nodes are the top-10 leaders in both networks (9 of which are shared by the two networks). The approximation uses the stretch method with 3 eigenpairs.}
\label{dolphin.btw}
\end{figure}

We conclude this section by computing exact and approximated betweenness centrality on a real network. The instance is a social network of dolphins (\textit{Tursiops truncatus}) belonging to a community that lives in the fjord of Doubtful Sound in New Zealand. The unusual conditions of this fjord, with relatively cool water and a layer of fresh water on the surface, have limited the departure of dolphins and the arrival of new individuals in the group, facilitating a strong social relationship within the dolphin community. The network is an undirected unweighted graph containing 62 nodes (dolphins) and 159 non-directional connections between pairs of dolphins. Two dolphins are joined by an edge if, during the observation period lasted from 1994 to 2001, they were spotted together more often than expected by chance. This network has been extensively studied by David Lusseau and co-authors, see, for instance, \citep{LN04}.

\begin{figure}[t]
\begin{center}
\includegraphics[scale=0.40, angle=-90]{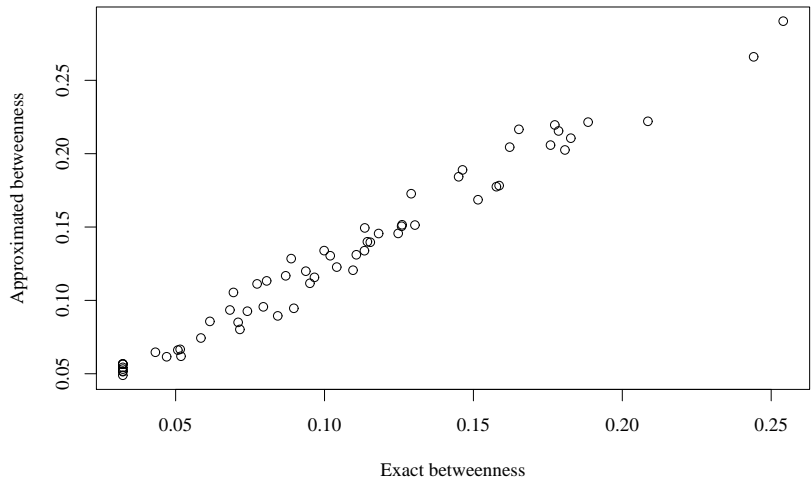}
\caption{A scatter plot of exact and approximated betweenness rankings: the rank of a dolphin in the exact ranking is plotted against the rank of the same dolphin in the approximated ranking. The approximation uses the stretch method with 3 eigenpairs.}
\label{fig.rankings}
\end{center}
\end{figure}

The  ranking obtained with the cutoff approximation method using only three eigenpairs of the graph Laplacian correlates at 0.92 with the exact ranking, and the mean change of rank between the two compilations is 5.4. Moreover, the stretch approximation method performs even better. Using the same number of eigenpairs (three), the approximated and exact rankings correlate at 0.99 and the mean change of rank between the two compilations is just 2 positions. Figure \ref{dolphin.btw} depicts the dolphin social network where the size of the node is proportional to its exact current-flow betweenness (graph on the left) and to its approximated current-flow betweenness (graph on the right). 
Moreover, Table \ref{tab.rankings} shows the top-10 of both compilations and Figure \ref{fig.rankings} gives the scatter plot of the two rankings. Nine dolphins over 10 are present in both top-10 rankings: the missing dolphins (DN63 and Grin) both rank 11th in the other ranking. Six dolphins have the same rank in both compilations (notably, the top-4 is the same). In fact, the stretch approximation method is already effective  using just one eigenpair (correlation at 0.98 and mean change at 2.9). These outcomes suggest that the proposed approximation methods for betweenness, in particular the stretch version, are effective. 

\begin{table}
\begin{center}
\begin{tabular*}{1\textwidth}{@{\extracolsep{\fill}}llll}
\textbf{Dolphin} &  \textbf{Exact} & \textbf{Dolphin} & \textbf{Approx} \\ \hline
Beescratch & 0.254 & Beescratch & 0.290 \\ \hline          
SN100      & 0.244 & SN100      & 0.266 \\ \hline       
Jet        & 0.209 & Jet        & 0.222 \\ \hline
SN9        & 0.189 & SN9        & 0.222 \\ \hline
Web        & 0.183 & SN4        & 0.220 \\ \hline
DN63       & 0.181 & Trigger    & 0.217 \\ \hline
Upbang     & 0.179 & Upbang     & 0.215 \\ \hline
SN4        & 0.177 & Web        & 0.211 \\ \hline
Kringel    & 0.176 & Kringel    & 0.206 \\ \hline
Trigger    & 0.165 & Grin       & 0.204 \\ \hline
\end{tabular*}
\end{center}
\caption{The top-10 rankings according to exact betweenness (column Exact) and approximated betweenness (column Approx). The approximation uses the stretch method with 3 eigenpairs.}
\label{tab.rankings}
\end{table}

%% file: experiments.tex
\section{Experimental evaluation} \label{experiments}

This section is devoted to the experimental evaluation of  effectiveness and efficiency of the proposed approximations for the generalized inverse of the Laplacian of a graph, and in particular for the  current-flow betweenness centrality scores of nodes. 

\subsection{Experimental setting}

In our experiments, we took advantage of the following two well-known network models. The first model is the random model, also known as Erd\H{o}s-R\'{e}nyi model (ER model for short) \citep{SR51,ER60}. According to this model, a network is generated by laying down a number of nodes and adding edges between them with independent probability for each node pair. This model generates a small world network with a binomial node degree distribution, which is peaked at a characteristic scale value. We will denote a graph drawn according to the ER model by $\mathrm{ER}(n, q)$, where $n$ is the number of nodes and $p = q/n$ is the edge probability. Such a graph has roughly $m = q (n-1) / 2$ edges. An ER graph is not necessarily connected, but it contains a giant connected component containing most of the nodes as soon as $q \geq 1$. We extracted the giant component from the generated graphs and used this component in our experiments.

However, many real networks are scale-free, that is, the degree distribution has a long-tail that roughly follows a power law \citep{N10}. A network model that captures the power law distribution of node degrees is known as cumulative advantage \citep{S55,P76}, which was later rediscovered and further investigated under the name preferential attachment or Barab\'{a}si-Albert model (BA model for short) \citep{BA99,BAJ99}. Such a model, as described by \citet{BA99}, has the following two main ingredients:

\begin{itemize}

\item \textit{Startup}: the graph has a single isolated node;

\item \textit{Growth}: additional nodes are added to the network one at a time;

\item \textit{Preferential attachment}: each node connects to the existing nodes with a fixed number $r$ of links. The probability that it will choose a given node is proportional to the number of links the chosen node already has.

\end{itemize}

The resulting network is a small-world graph with a power law degree distribution: most of the nodes (the trivial many) will have low degree and a small but significant share of nodes (the vital few or hubs) will have an extraordinary high degree. We will denote a graph drawn according to the BA model as $\mathrm{BA}(n, r)$, where $n$ is the number of nodes and $r$ is the number of edges attached to each node added during the preferential attachment step. Such a graph has  $m = r (n-1)$ edges. Notice that the resulting graph is connected but it is not necessarily simple: multiple edges might exist between the same pair of nodes. We simplified the graph in our experiments, removing the multiple edges, if any.

In this section we give results for unweighted networks only. We also experimented with weighted graphs with edge weights drawn from a random uniform distribution, but we noticed no particular discrepancy with respect to the unweighted case.  

The experiments have been performed within the R computing environment, taking advantage of the \textit{igraph} package for the generation of networks. We used of R interface to LAPACK (Linear Algebra Package)  \citep{ABBBDDDGHMS99} for matrix inversion and eigenpairs generation, as well as the R interface to ARPACK (Arnoldi Package) \citep{LSY98}, when few eigenpairs of large sparse matrices are necessary.  Moreover, we exploited the C programming language for a faster implementation of the computation of betweenness scores using Equation \ref{cent}, calling the C compiled code from the R environment  (R is rather slow at iterative algorithms that require loops iterated many times). All experiments are run on a MacBook Pro equipped with a 2.3 GHz Intel Core i5 processor, 4 GB 1333 MHz DDR3 memory, running Mac OS X Lion 10.7.2. 

\subsection{Distribution of eigenvalues and of node degrees}

We begin the experimental investigation with an exploratory analysis of the distribution of eigenvalues of the Laplacian, comparing with the distribution of node degrees on both ER and BA networks. This study is interesting to understand the behavior of the proposed approximations. 

The exploratory experiment uses graphs $\mathrm{ER}(n = 1000, q = 10)$ and $\mathrm{BA}(n = 1000, r = 5)$. Notice that the used graphs have the same number of nodes and approximately the same number of edges, hence a similar edge density. Nevertheless, they differ in the way the edges are placed among nodes. 

\begin{figure}[t]
\begin{center}
\includegraphics[scale=0.20, angle=-90]{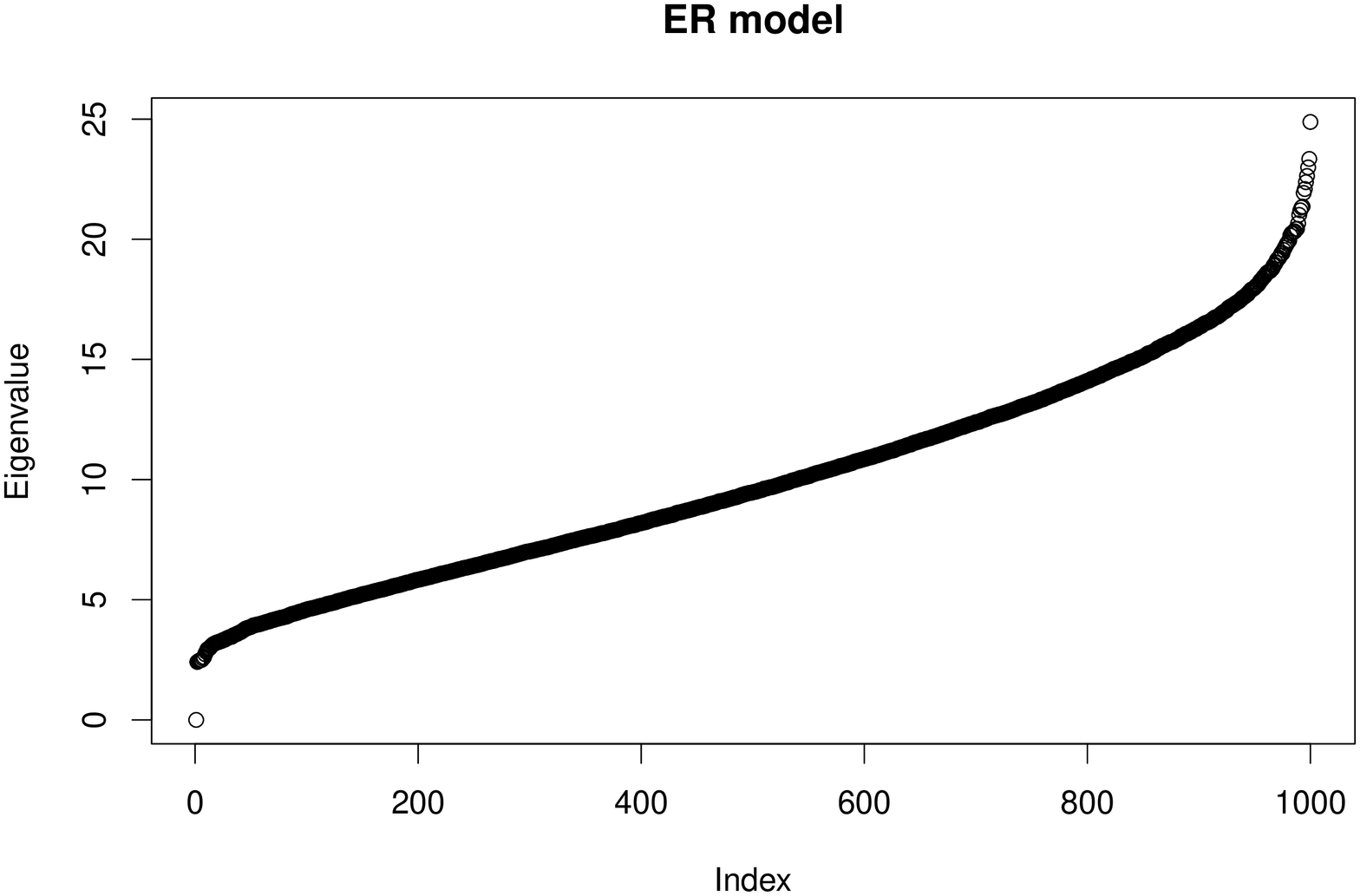}
\includegraphics[scale=0.20, angle=-90]{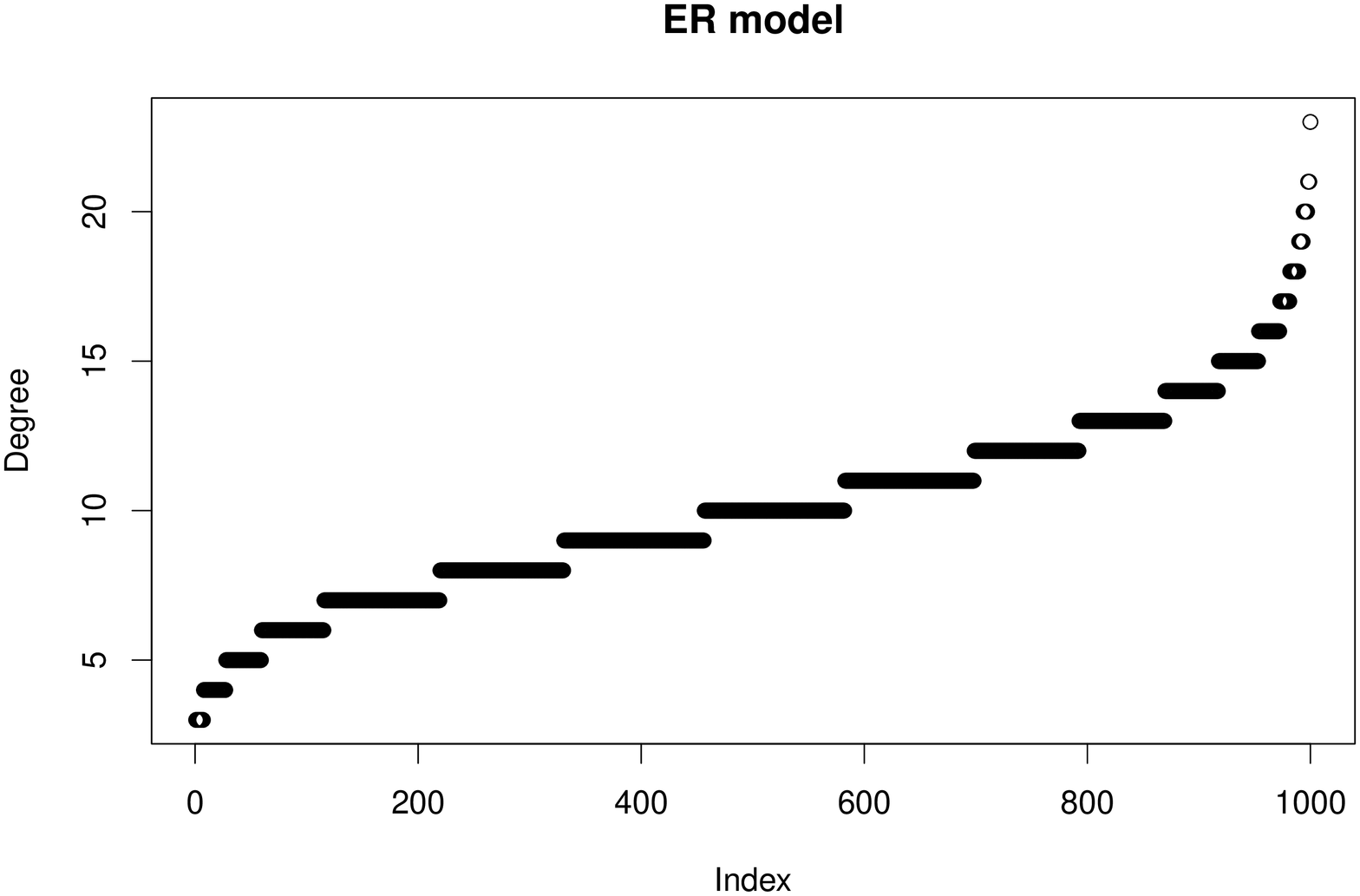}
\caption{Plot of sorted eigenvalues (left plot) and of sorted degrees (right plot) on $\mathrm{ER}(n = 1000, q = 10)$}
\label{0.E3}
\end{center}
\end{figure}

\begin{figure}[t]
\begin{center}
\includegraphics[scale=0.20, angle=-90]{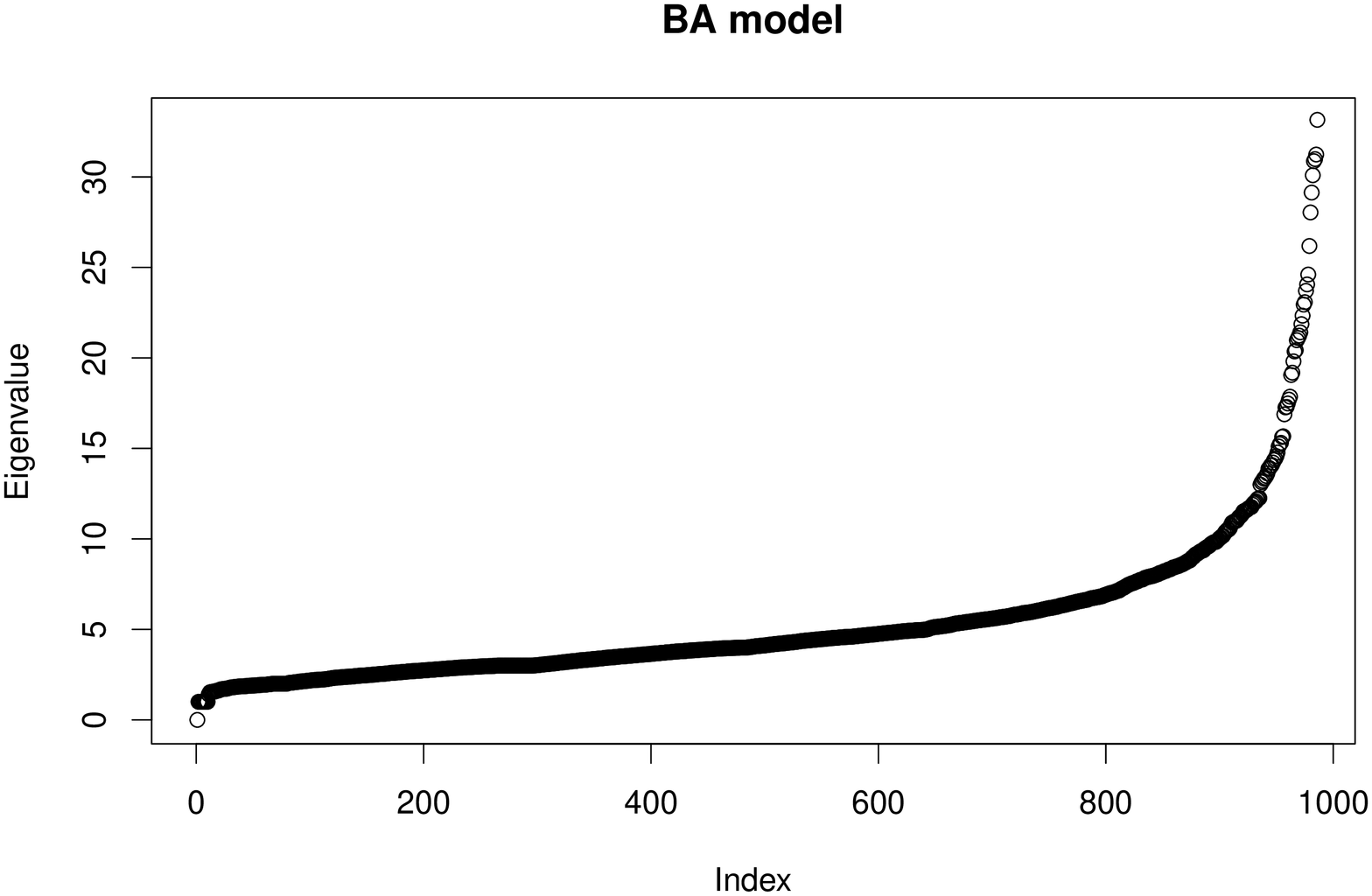}
\includegraphics[scale=0.20, angle=-90]{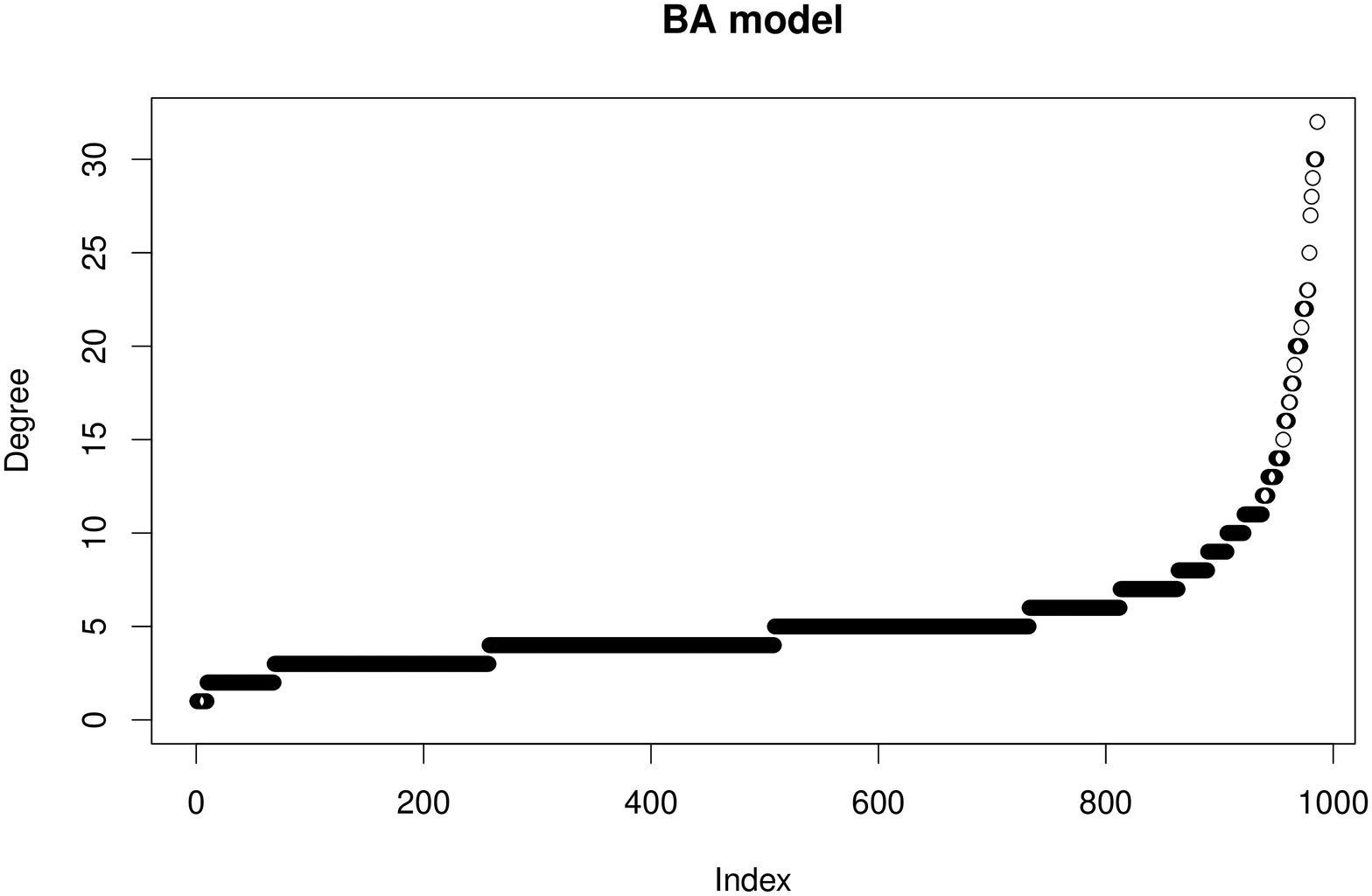}
\caption{Plot of sorted eigenvalues (left plot, up to index 980) and of of sorted node degrees (right plot, up to index 980)  on $\mathrm{BA}(n = 1000, r = 5)$}
\label{0.F3}
\end{center}
\end{figure}

\begin{figure}[t]
\begin{center}
\includegraphics[scale=0.20, angle=-90]{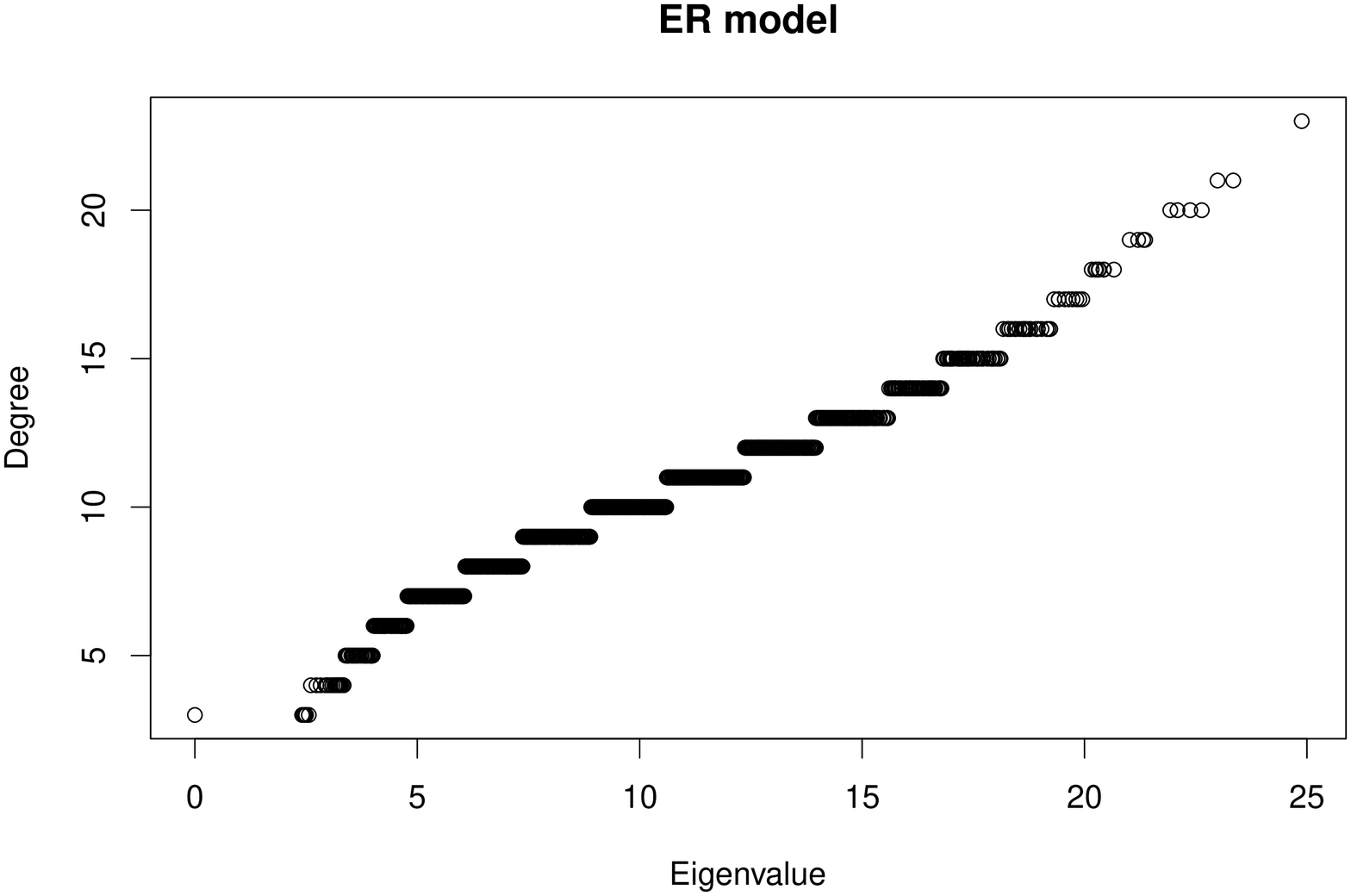}
\includegraphics[scale=0.20, angle=-90]{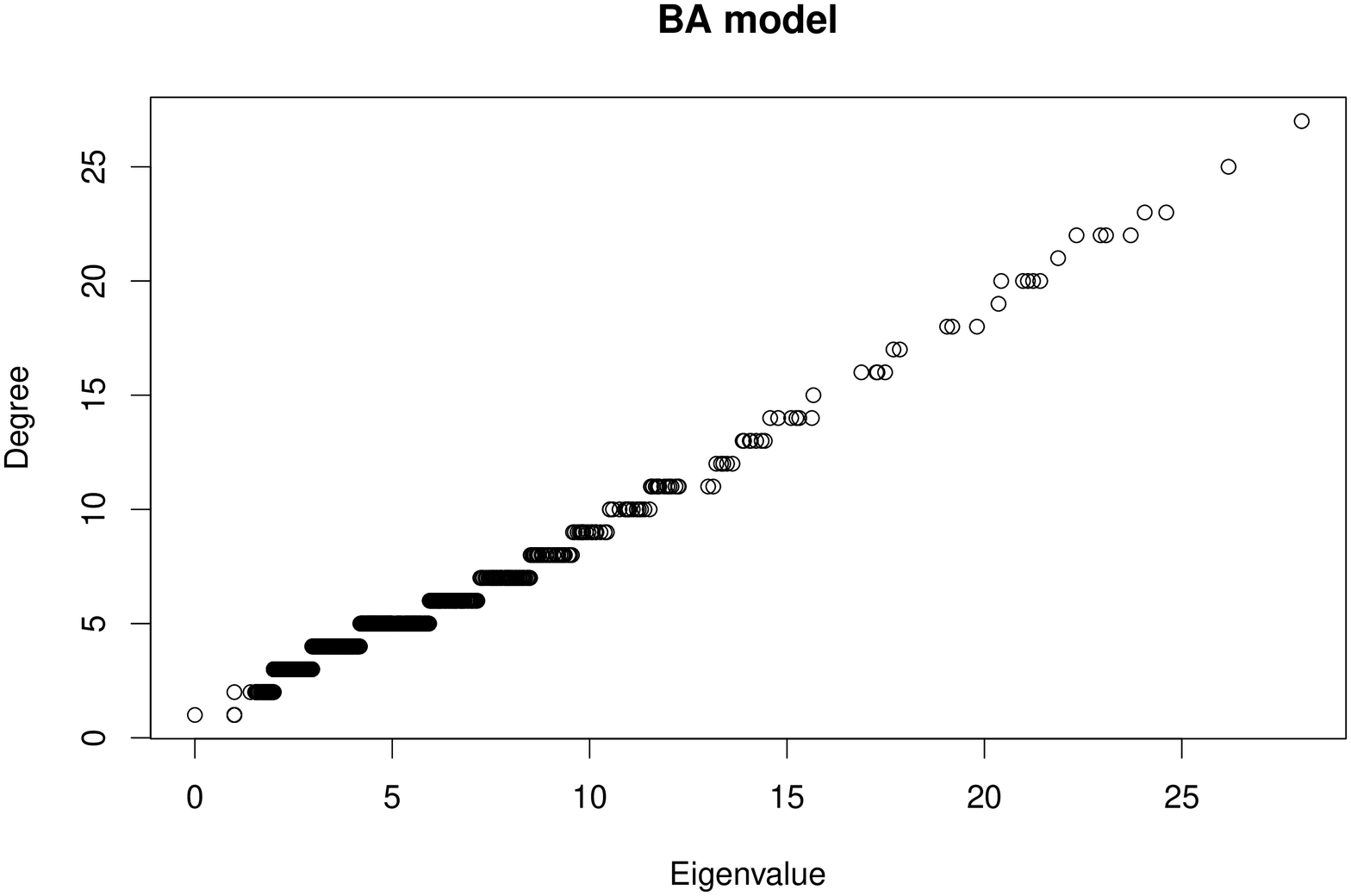}
\caption{Scatter plot of eigenvalues versus node degrees on $\mathrm{ER}(n = 1000, q = 10)$ (left plot) and on $\mathrm{BA}(n = 1000, r = 5)$ (right plot)}
\label{0.E5}
\end{center}
\end{figure}

\begin{figure}[t]
\begin{center}
\includegraphics[scale=0.20, angle=-90]{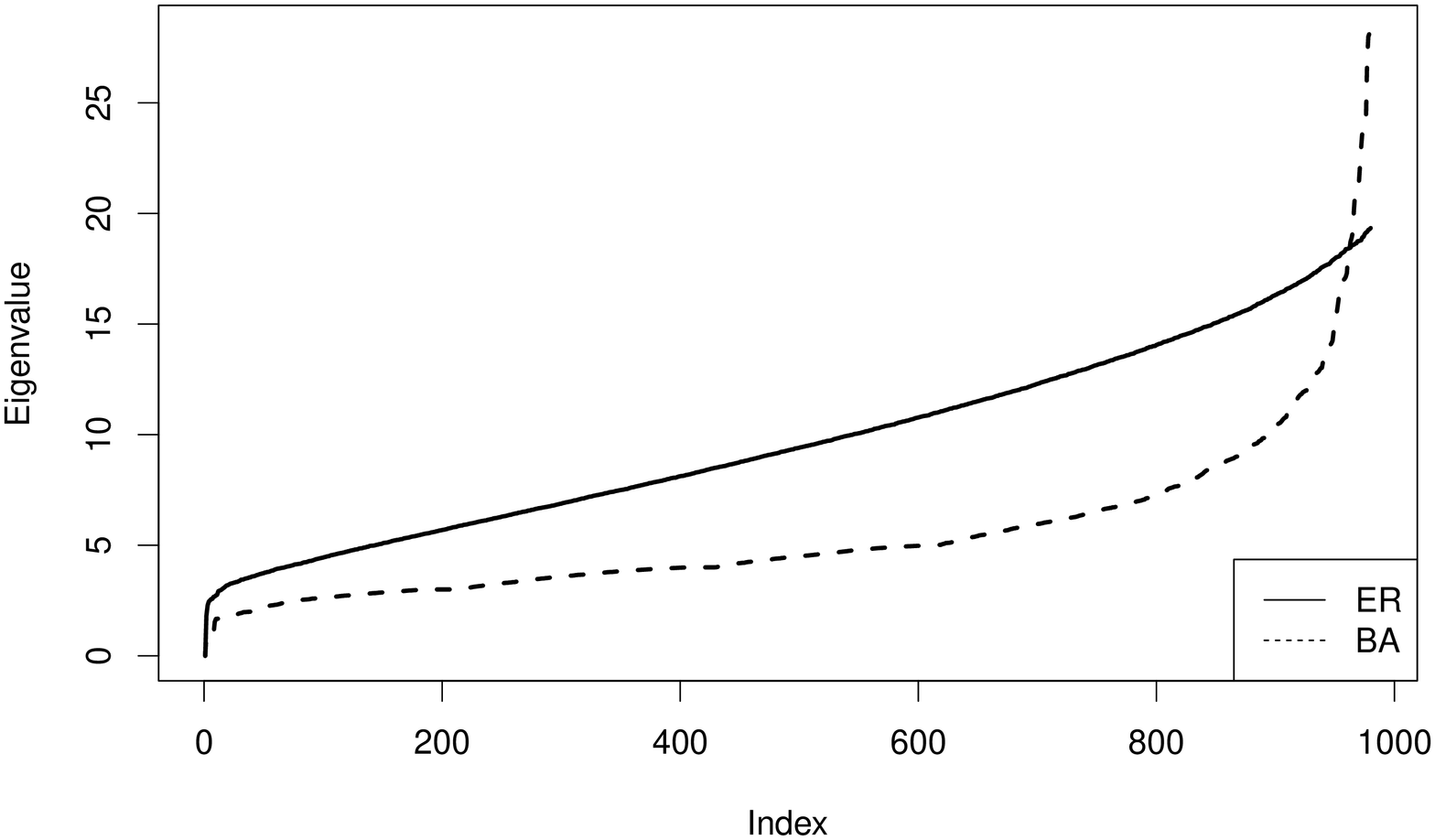}
\includegraphics[scale=0.20, angle=-90]{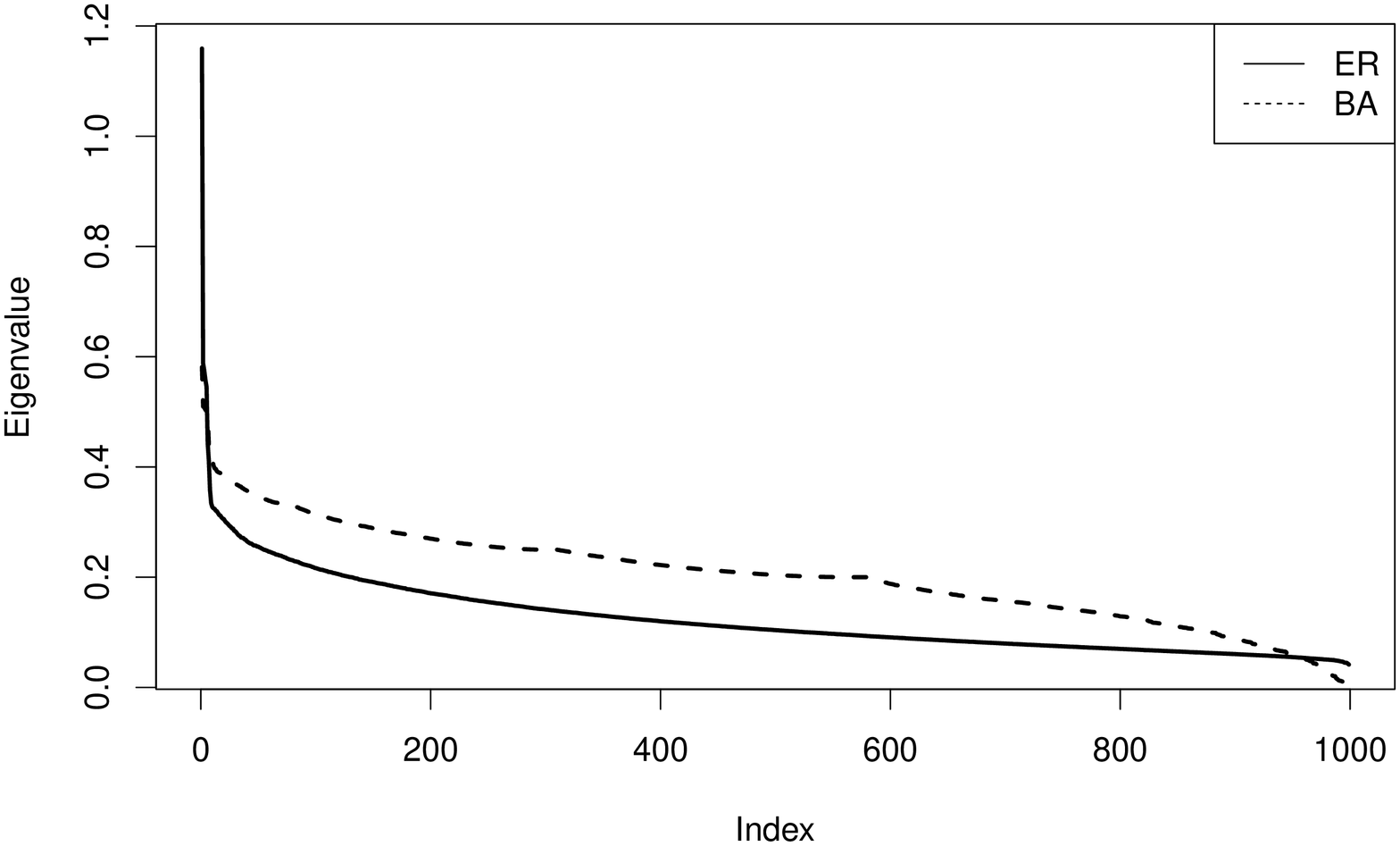}
\caption{Plot of eigenvalues of Laplacian (left plot, up to index 980) and of generalized inverse Laplacian (right plot, up to index 980) on $\mathrm{ER}(n = 1000, q = 10)$ and $\mathrm{BA}(n = 1000, r = 5)$.}
\label{0.EF}
\end{center}
\end{figure}

Figures~\ref{0.E3} and \ref{0.F3} explore the distribution of eigenvalues of the Laplacian as well as the distribution of node degrees on ER and BA graphs, respectively. We notice that the eigenvalues of the Laplacian are well approximated by the degrees of the nodes, an interesting phenomenon already investigated in \citet{ZCY10}. For the ER model, eigenvalues and degrees have a correlation of 0.989; the Euclidean distance between the two sorted vectors, relative to the norm-2 of the eigenvalue vector, is 0.13.  Using a BA model, eigenvalues and degrees are correlated at 0.983 and the relative Euclidean distance between the two sorted vectors is 0.02. Figure~\ref{0.E5} shows a scatter plot of  Laplacian eigenvalues versus node degrees.

The plots of the sorted eigenvalues of the Laplacian on ER and BA models are visibly different (Figure~\ref{0.EF}, left plot). In the initial part (up to index 800), ER eigenvalues grow faster than BA eigenvalues; however, in the tail of the plot, BA eigenvalues rapidly catch up, overtaking ER eigenvalues around index 960. From here to the end of the sequence, BA eigenvalues literally explode (notice that, in order to appreciate the behavior of smaller eigenvalues, the figure shows the eigenvalue curve up to index 980). The eigenvalues of the generalized inverse of the Laplacian, which are the inverse of the Laplacian eigenvalues, are shown in Figure~\ref{0.EF}, right plot. Both curves decrease rapidly in the beginning, but the ER curve has a more significant drop; after this initial fall, both curves decreases gently, although the BA line declines faster. This might be a hint of the effectiveness of the stretch approximation method: the tail of the eigenvalue curve has a small slope, hence the eigenvalues of the tail can be well approximated by a single middle value. 

\subsection{Effectiveness of approximations}

\begin{figure}[t]
\begin{center}
\includegraphics[scale=0.20, angle=-90]{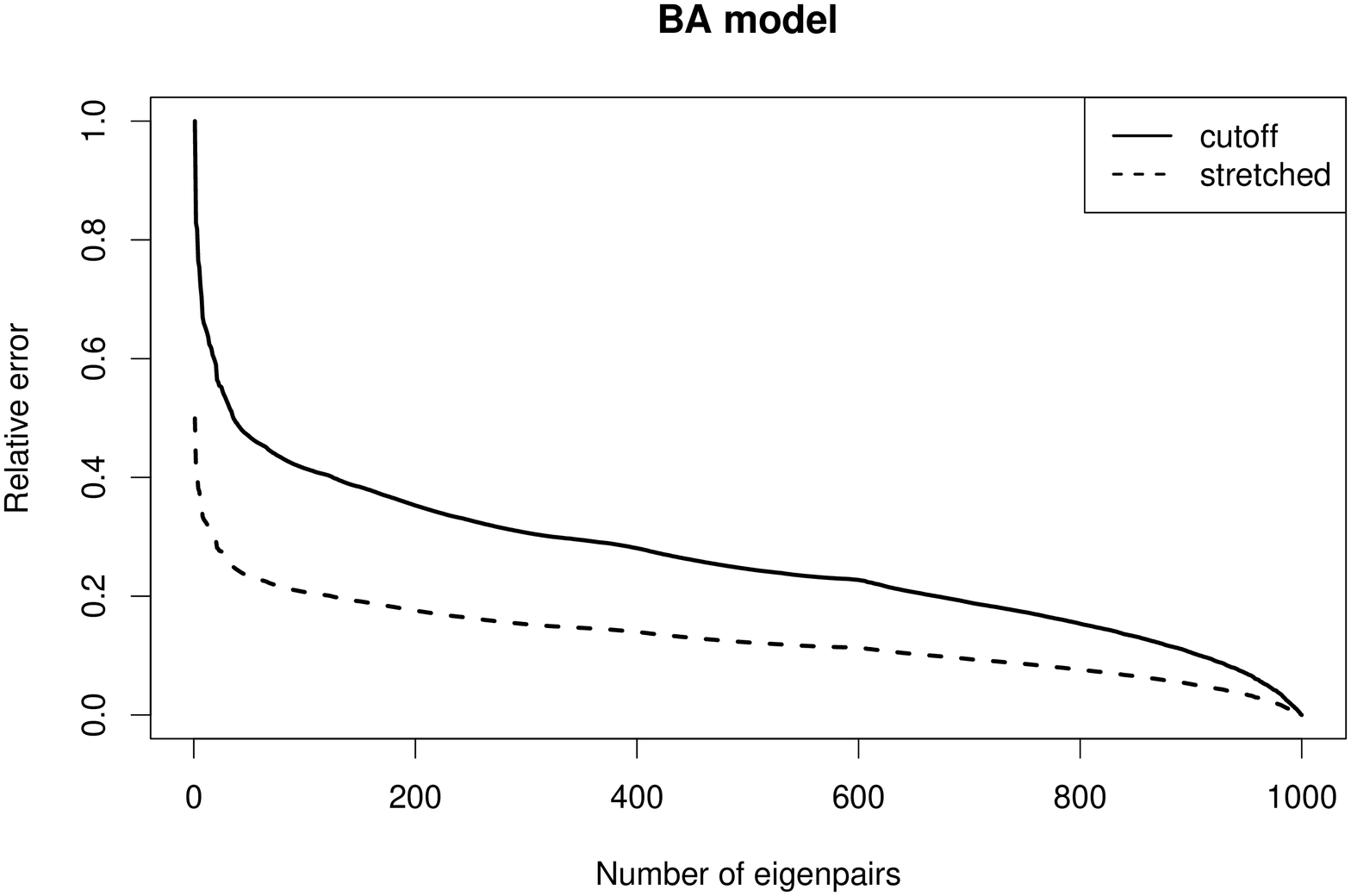}
\includegraphics[scale=0.20, angle=-90]{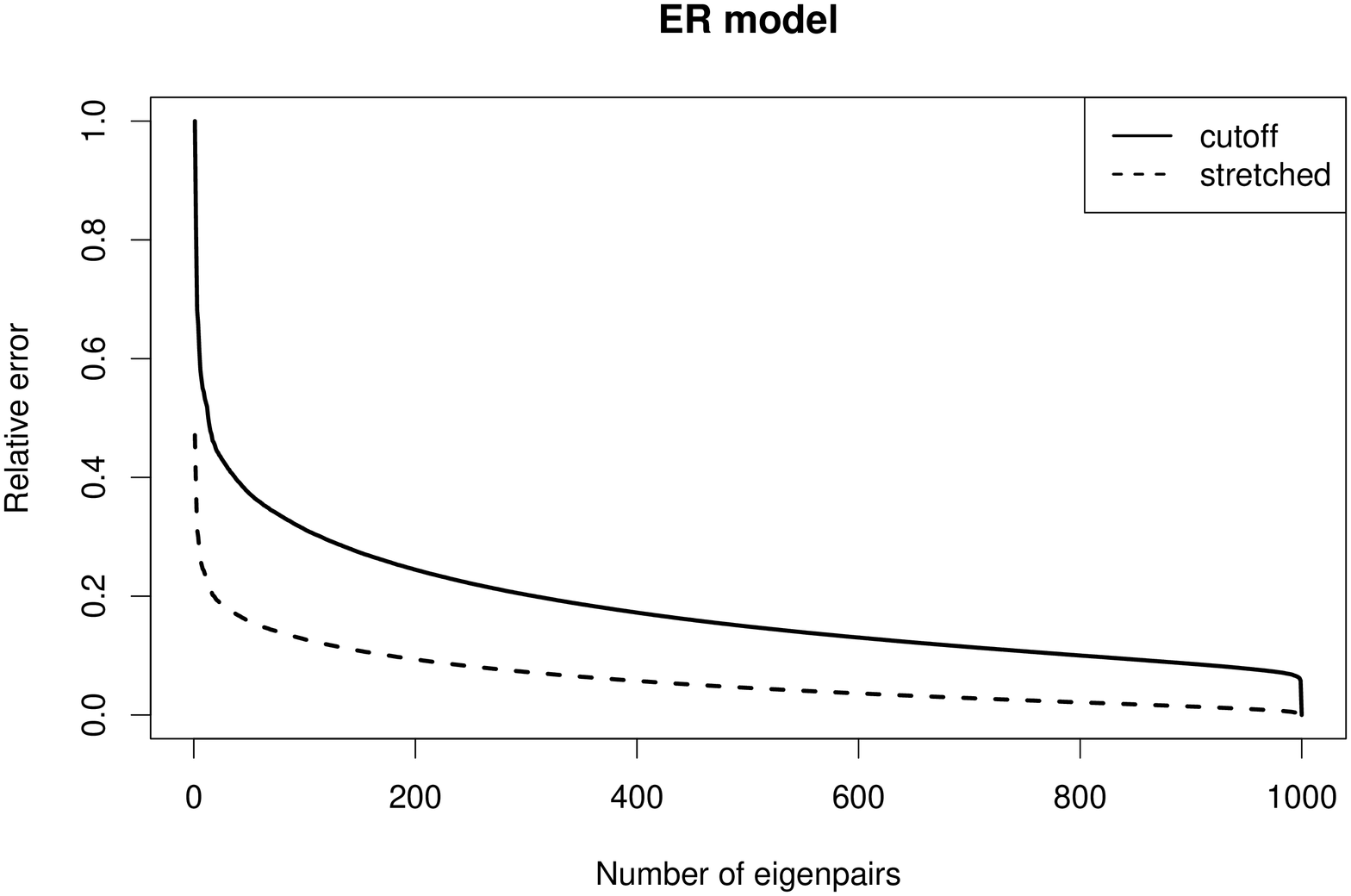}
\caption{Relative 2-norm error of cutoff and stretch approximations on $\mathrm{BA}(n = 1000, r = 5)$ (left plot) and $\mathrm{ER}(n = 1000, q = 10)$ (right plot).}
\label{0.23}
\end{center}
\end{figure}

\begin{figure}[t]
\begin{center}
\includegraphics[scale=0.20, angle=-90]{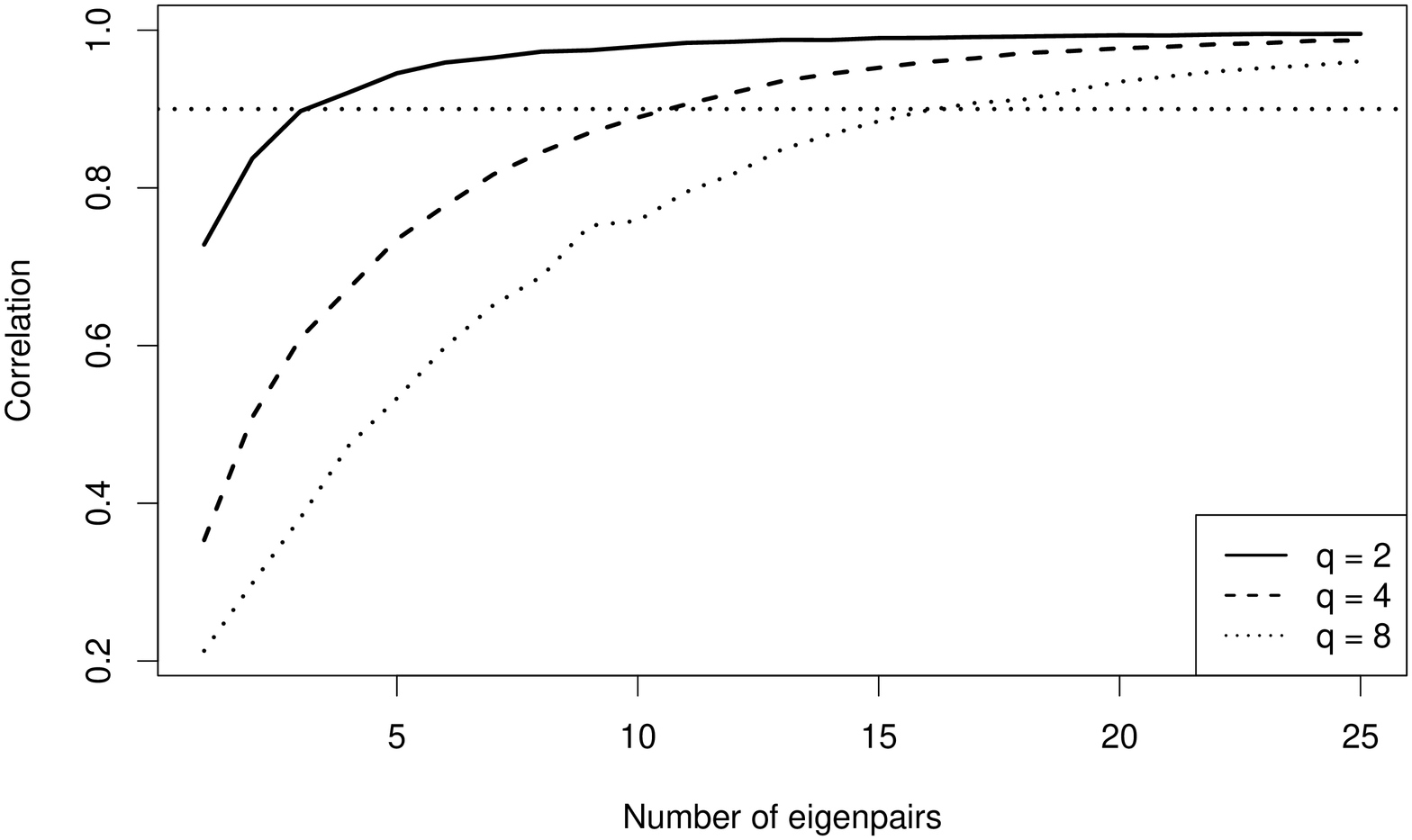}
\includegraphics[scale=0.20, angle=-90]{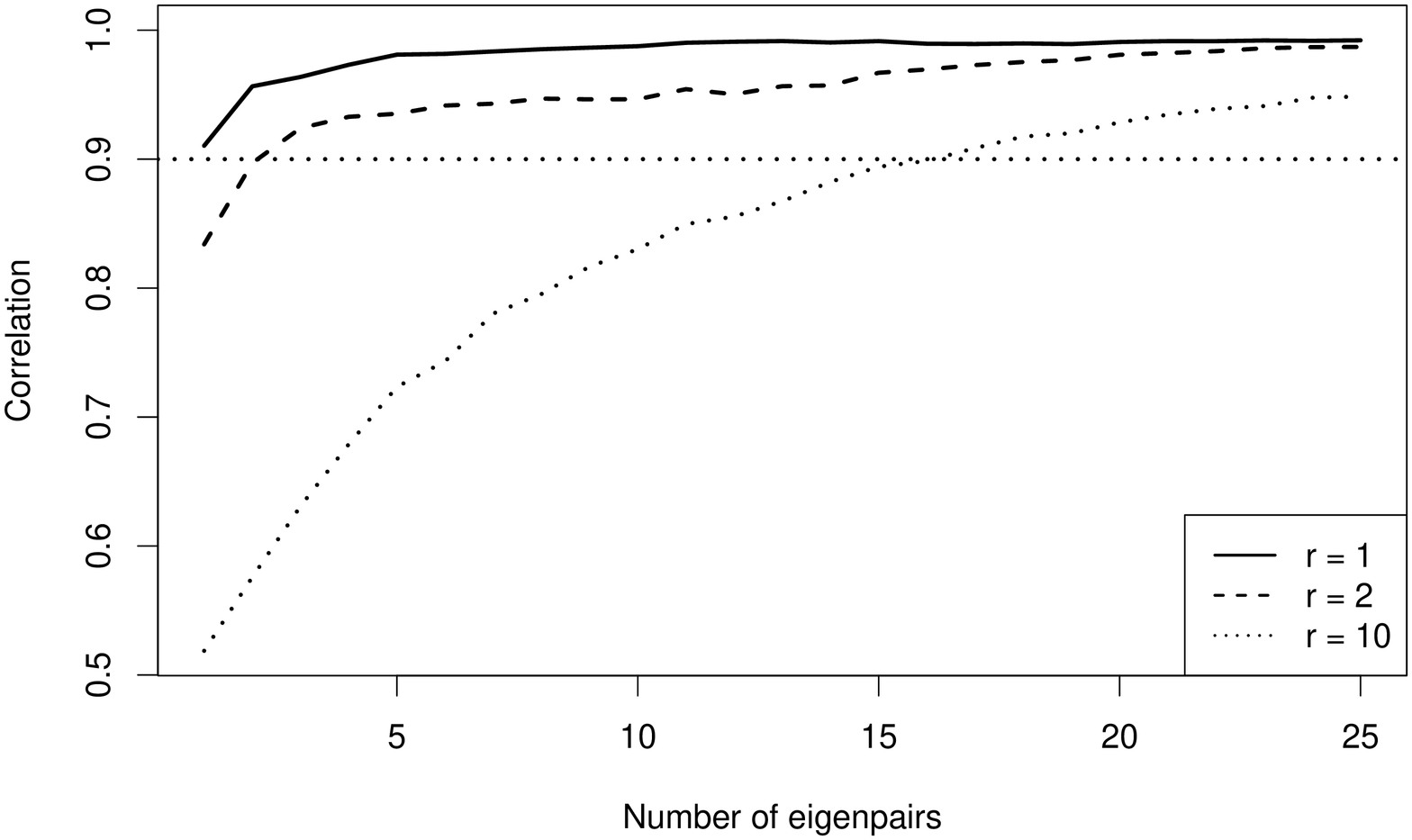}
\caption{Effectiveness of cutoff approximation for betweenness increasing the number of eigenpairs from 1 to 25 on $\mathrm{ER}(n = 100, q)$ for different values of $q$ (plot on the left) and on $\mathrm{BA}(n = 100, r)$ for different values of $r$ (plot on the right).}
\label{e1}
\end{center}
\end{figure}

\begin{figure}[t]
\begin{center}
\includegraphics[scale=0.40, angle=-90]{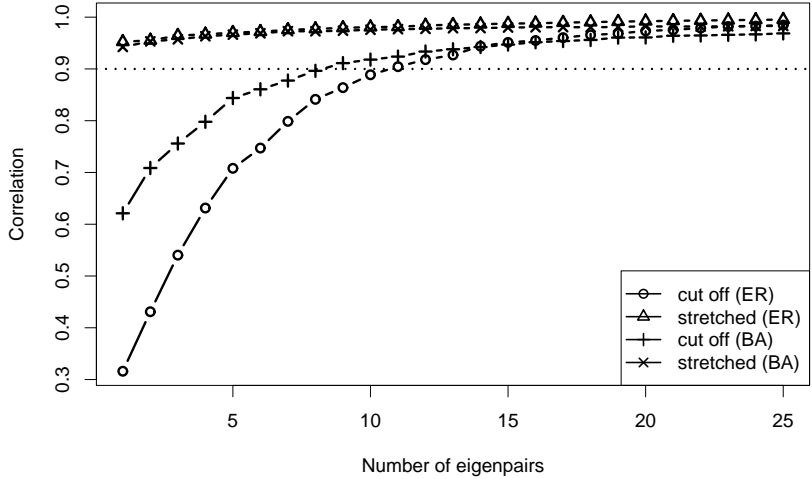}
\caption{Comparison of cutoff and stretch approximations for betweenness increasing the number of eigenpairs from 1 to 25 on $\mathrm{ER}(n = 100, q=4)$ and $\mathrm{BA}(n = 100, r=2)$.}
\label{e3}
\end{center}
\end{figure}

\begin{figure}[t]
\begin{center}
\includegraphics[scale=0.40, angle=-90]{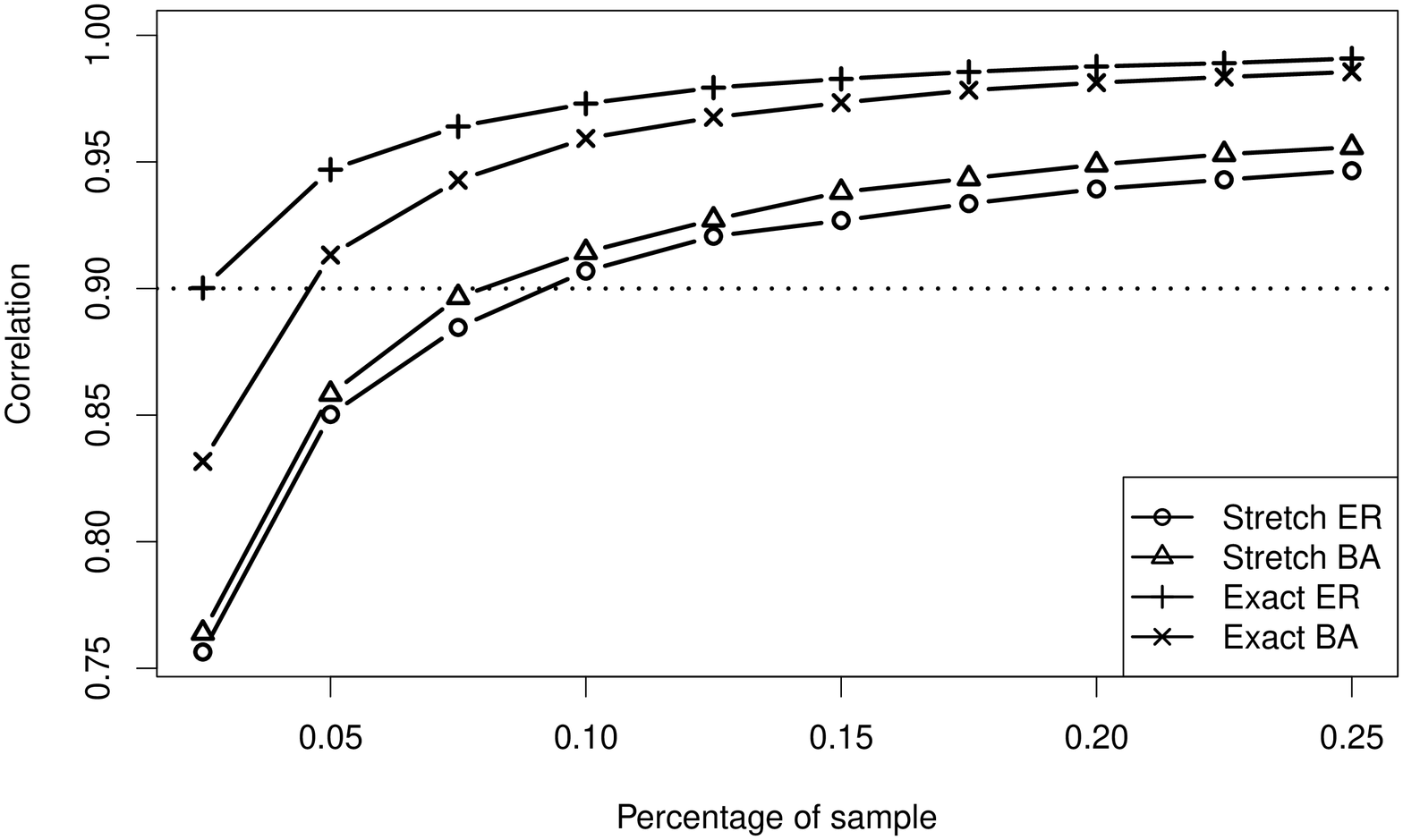}
\caption{Comparison of the probabilistic approach on both exact and stretch betweenness (one eigenpair) increasing the size of the node pair sample from 2.5\% to 25\% on $\mathrm{ER}(n = 100, q = 4)$ and $\mathrm{BA}(n = 100, r = 2)$.}
\label{e17}
\end{center}
\end{figure}

In this section we investigate the effectiveness of approximations of the generalized inverse of the Laplacian and, in particular, of the current-flow betweenness rankings. We first study the relative 2-norm error of the $k$-th cutoff and stretch approximations, defined as the ratio between 
$\|G^+-T^{(k)}\|_2 $ and $\|G^+\|_2$ (cutoff approximation error), and the ratio between $\|G^+-S^{(k)}\|_2$ and $\|G^+\|_2$ (stretch approximation error). Figure \ref{0.23} shows the approximation error as the number of eigenpairs $k$ grows from $1$ to $n$. On both network models, the stretch approximation is significantly more effective than the cutoff one: on average, the stretch approximation error is 30\% and 50\% of the cutoff approximation error on ER and BA graphs, respectively. In particular, the stretch approximation is more effective on ER graphs, as predicted by the spectral behavior highlighted in Figure~\ref{0.EF}.

We next test the effectiveness of the approximations applied to current-flow betweenness.
The quality of the approximations is establishing by computing the correlation coefficient among the approximated and exact node rankings. We used the Pearson product-moment correlations, where the input data is logarithmically transformed when it is not normally distributed. This coefficient is largely used in sciences to measure the similarity of two rankings; it runs from $-1$ to $1$ where values close to $-1$ indicate negative correlation (one ranking is the reverse of the other), values close to $0$ correspond to null correlation (the two rankings are independent), and values close to $1$ denote positive correlation (the two rankings are similar). A good approximation, in our assessment, is a method that approaches the exact ranking with correlation of at least 0.9. 

The effectiveness experiments are run on ER and BA graphs with 100 nodes and an increasing number of edges: we used $q=2,4,8$ for the the ER model and $r=1,2,10$ for the BA model. The corresponding ER and BA graphs have roughly the same density. Notice that a BA graph with $r=1$ is a tree, hence an acyclic graph, while $r > 1$ generates graphs with loops. We always generated a sample of 100 such graphs and took the average of the observed correlation coefficients. 

Figure~\ref{e1} shows the effectiveness of the cutoff approximation for betweenness on ER and BA graphs with increasing density. The quality of the approximation increases with the number of eigenpairs that are used in the approximation and decreases, for both network models, as the graphs become more dense. A cumulative comparison between cutoff and stretch approximations as well as among ER and BA network models is given in Figures~\ref{e3}. The stretch approximation performs neatly better than the cutoff one. Regardless of the graph model and of the graph density, the effectiveness of the stretch method is well above the quality threshold of 0.9 already with a single eigenpair. 

Finally, we tested the probabilistic approximation of betweenness that uses only a sample of node pairs for solving Equation \ref{cent}. In Figures~\ref{e17} we compare the probabilistic approach on both the exact version of betweenness, that uses the full generalized Laplacian inverse matrix for the computation of betweenness scores, and the stretch version of betweenness, that uses the stretch approximation method with one eigenpair only. The plot shows the effectiveness of the combined approximation algorithm that uses the stretch method (with a single eigenpair) for the approximation of the generalized Laplacian inverse and the probabilistic approach for the computation of betweenness scores: less than 10\% of the node pair space is sufficient to get an approximated ranking that correlates at 0.9 with the exact ranking.

\subsection{Efficiency of approximations}

\begin{figure}[t]
\begin{center}
\includegraphics[scale=0.20, angle=-90]{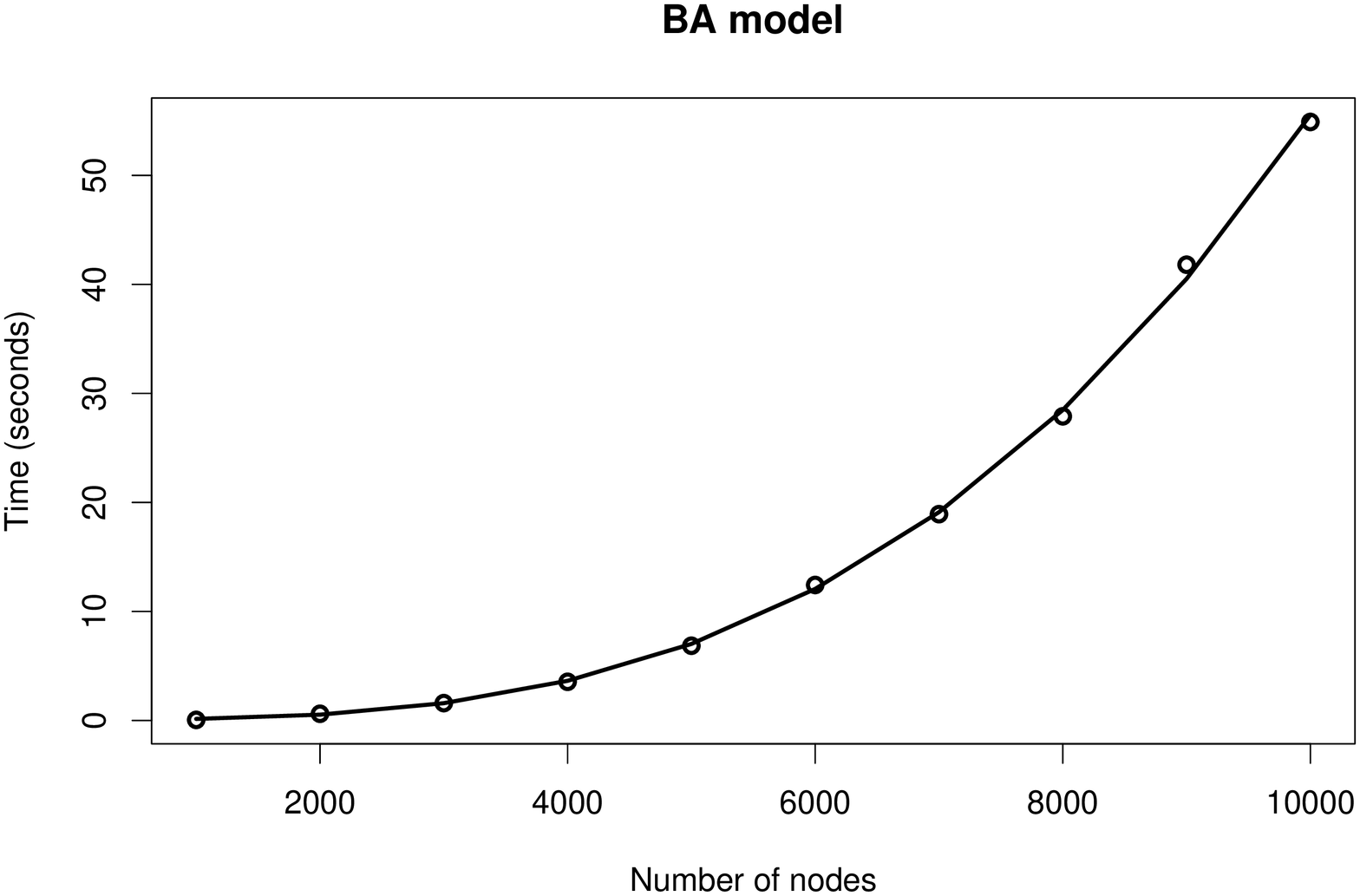}
\includegraphics[scale=0.20, angle=-90]{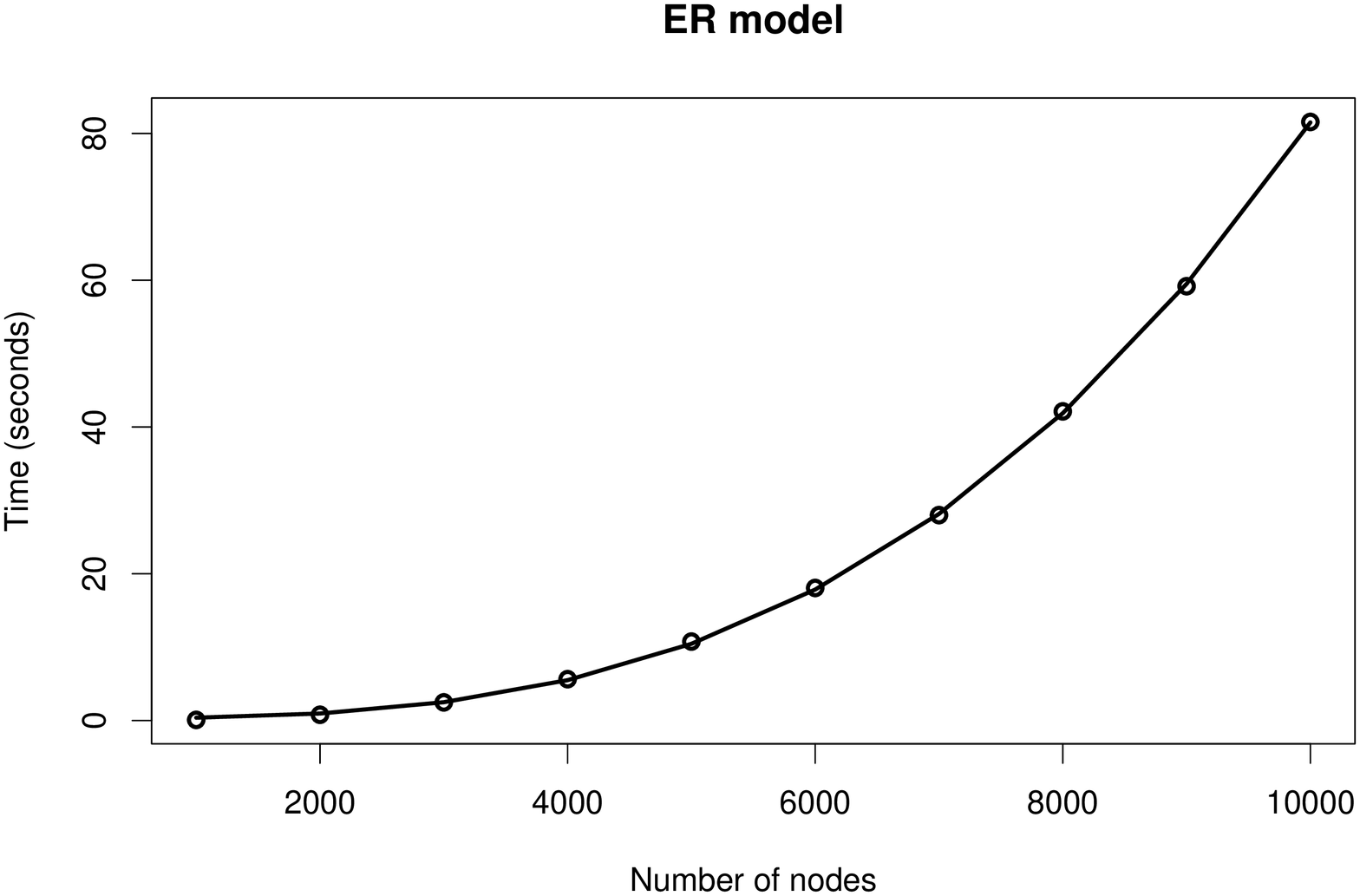}
\caption{Efficiency of the computation of the generalized inverse of the Laplacian matrix of BA graphs ($r=2$, plot on the left) and ER graphs ($q = 4$, plot on the right) with increasing number of nodes. The data fits a cubic curve $ax^3 + b$ with $b=8.906 \cdot 10^{-2}$ and $a=5.545 \cdot 10^{-11}$ for the BA model (R-Square = 0.9992), and a cubic curve $ax^3 + b$ with $b=3.038 \cdot 10^{-1}$ and $a=8.120 \cdot 10^{-11}$ for the ER model (R-Square = 0.9999).} 
\label{e22a}
\end{center}
\end{figure}

\begin{figure}[t]
\begin{center}
\includegraphics[scale=0.20, angle=-90]{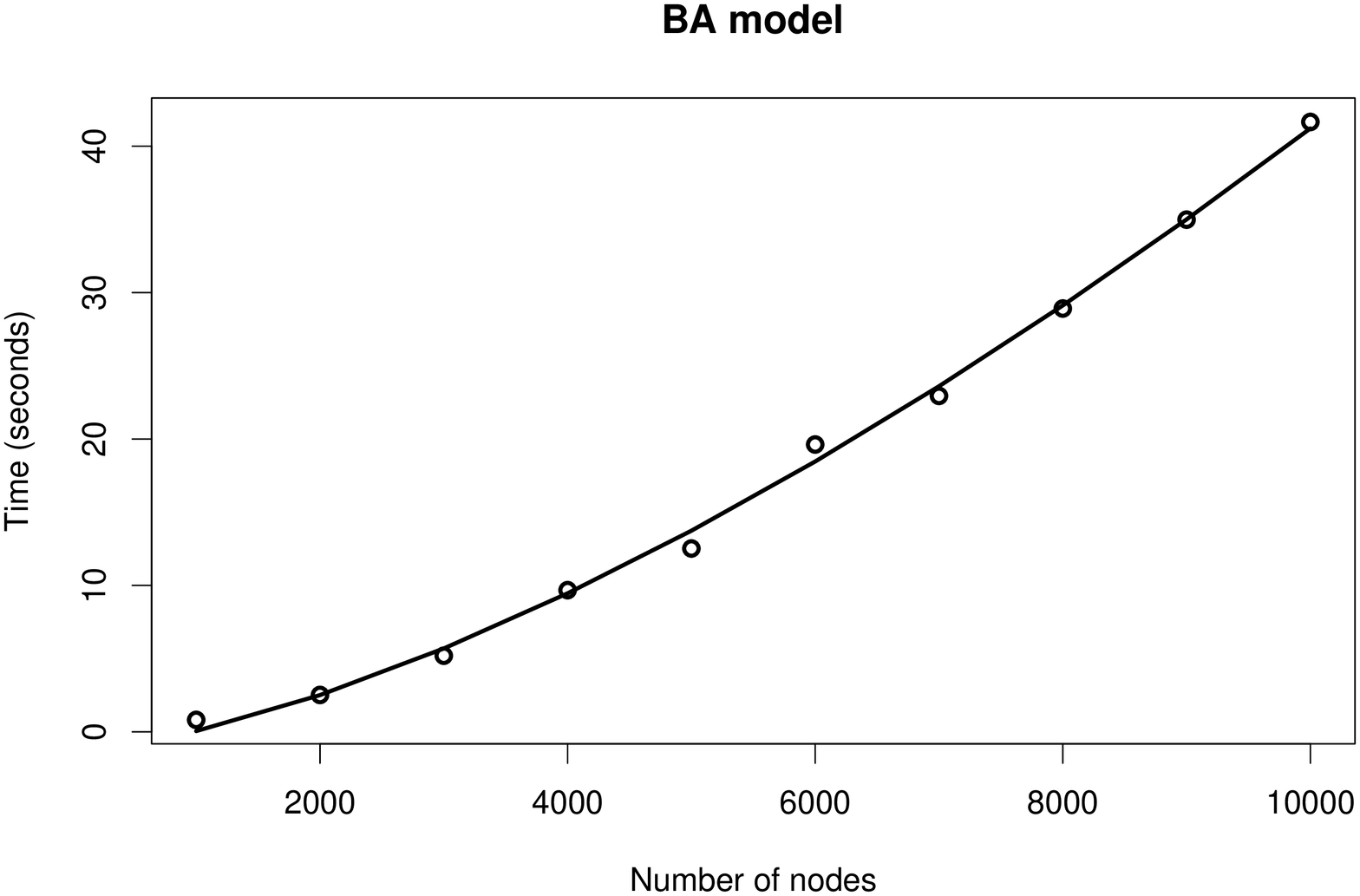}
\includegraphics[scale=0.20, angle=-90]{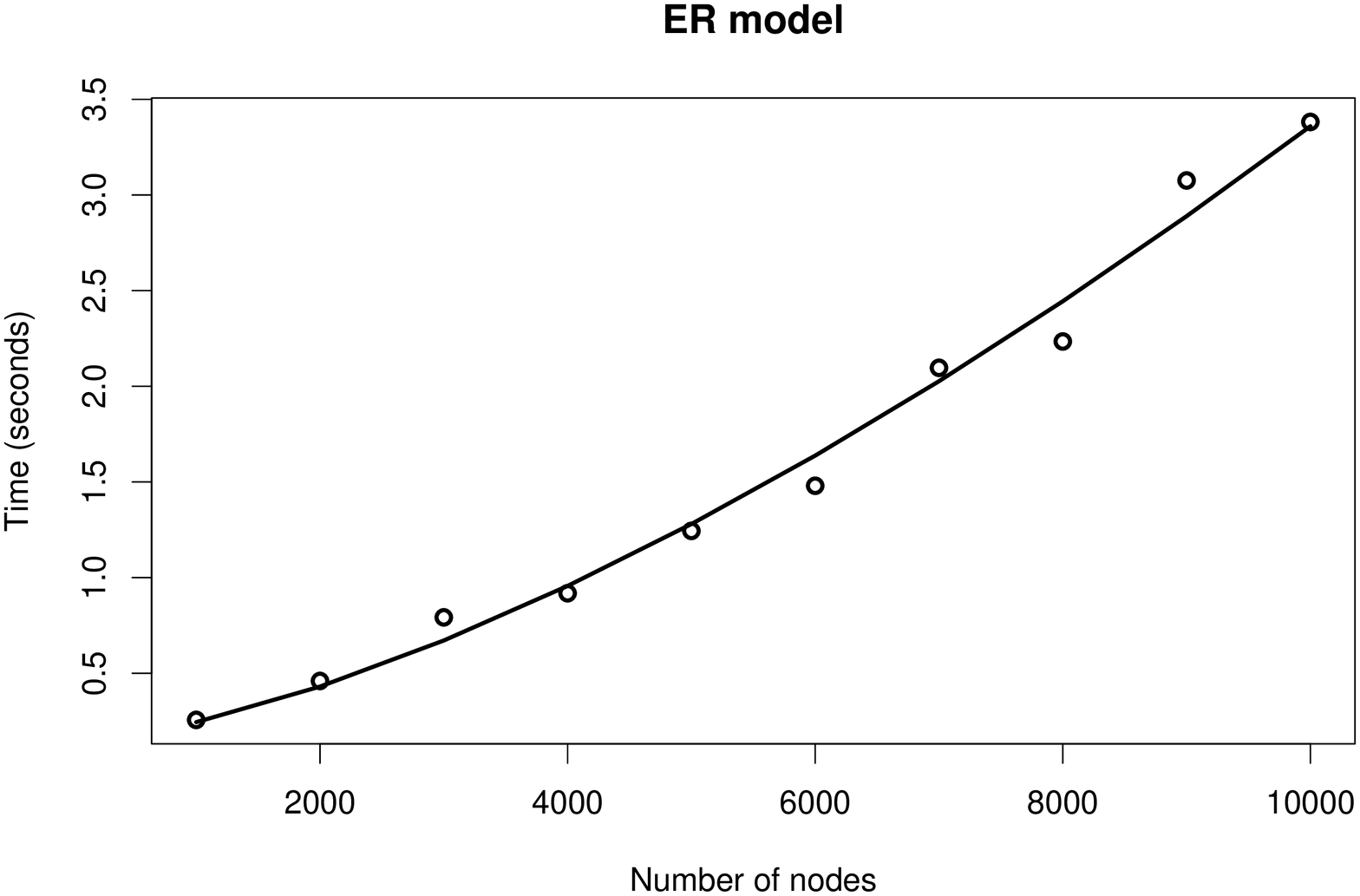}
\caption{Efficiency of the computation of the second smallest eigenpair of the Laplacian matrix of BA graphs ($r=2$, plot on the left) and ER graphs ($q = 4$, plot on the right) with increasing number of nodes. The data fits a curve $ax^{1.5} + b$ with $b=-1.292$ and $a=4.25 \cdot 10^{-5}$ for the BA model (R-Square = 0.9976), and a curve $ax^{1.5} + b$ with $b=1.426 \cdot 10^{-1}$ and $a=3.217 \cdot 10^{-6}$ for the ER model (R-Square = 0.9877).}
\label{e21b}
\end{center}
\end{figure}

In this section we show the outcomes of experiments about the efficiency of the proposed approximations for the generalized inverse of the Laplacian matrix. We compare elapsed time of the computation of the full generalized inverse of the Laplacian with that of the computation of the approximation that uses only one eigenpair (the second smallest eigenvalue and the corresponding eigenvector).

Figure~\ref{e22a} shows the elapsed time necessary for the computation of the generalized Laplacian inverse for BA and ER networks. The time growth is $O(n^3)$ and the memory requirement is $O(n^2)$ on both models, where $n$ is the number of nodes of the graph. These costs make the computation not feasible on large networks. For instance, for a large network with $10^6$ nodes, the necessary time would be 642 days with a storage of about one terabyte. On the other hand, Figure~\ref{e21b} gives the elapsed time necessary for the computation of  the approximation that uses only one eigenpair for BA and ER networks. For both models, the time growth is $O(n^{1.5})$ and the memory requirement is $O(n)$. Notice that the computation on ER graphs is one order of magnitude more efficient compared to BA graphs. These complexity make it possible to approximate the generalized inverse of the Laplacian, and in particular current-flow betweenness scores, on relatively large networks: on a standard machine, the computation on a random network with $10^6$ nodes would last less than one hour and it would take 
about 12 hours (a reasonable time) on a scale-free network with the same number of nodes. In both cases the required storage is only one megabyte.

%% file: conclusion.tex
\section{Conclusion} \label{conclusion}

We have proposed effective and efficient methods for approximating the generalized inverse Laplacian matrix of a graph, a matrix that arises in many graph-theoretic applications. A notable such application investigated in the present contribution is current-flow betweenness centrality, a measure for finding central nodes in a graph that lie between many other nodes. Our approximation methods turn out to be suitable when the network at hand is large, so that computing the full generalized inverse is not realistic. 

Many other applications can be sought, for instance, the approximation of resistance distance matrix of a graph. Resistance distance is a metric on the graph, alternative to the classical shortest-path distance, that was defined, independently, by \citet{SZ89}, following an information-theoretic approach, and by \citet{KR93}, following an electrical-theoretic approach. The resistance distance notion has different interesting interpretations and many applications, even outside of network science \citep{GBS08}. It turns out that resistance distance matrix can be immediately defined in terms of the generalized inverse Laplacian matrix \citep{GBS08}, allowing us to extend our approximation methods to resistance distance matrix as well.